\documentclass[useAMS,fleqn,usenatbib]{mnras}
\usepackage[utf8]{inputenc}
\usepackage{amssymb,amsmath}
\usepackage{graphicx}
\usepackage[dvipsnames]{xcolor}
\hypersetup{colorlinks=true,allcolors=teal}
\usepackage[flushleft]{threeparttable}

\newcommand{\be}{\begin{equation}}
\newcommand{\ee}{\end{equation}}
\def\bear#1\ear{\begin{align}#1\end{align}}
\newcommand{\nline}{\notag \\}

\newcommand{\e}{\mathrm{e}}

\newcommand{\Msun}{\mathrm{M}_{\odot}}

\newcommand{\eqn}[1]{equation~(\ref{#1})}

\newcommand{\secn}[1]{Section~\ref{#1}}
\newcommand{\appndx}[1]{Appendix~\ref{#1}}
\newcommand{\fig}[1]{Fig.~\ref{#1}}
\newcommand{\figs}[1]{Figs.~\ref{#1}}
\newcommand{\tab}[1]{Table~\ref{#1}}

\usepackage{newtxtext,newtxmath}

\usepackage[T1]{fontenc}

\DeclareRobustCommand{\VAN}[3]{#2}
\let\VANthebibliography\thebibliography
\def\thebibliography{\DeclareRobustCommand{\VAN}[3]{##3}\VANthebibliography}

\usepackage{graphicx}	
\usepackage{amsmath}	
\usepackage{subfig}

\title[21~cm power spectra using SCRIPT]{Efficient exploration of reionization parameters for the upcoming 21~cm observations using a photon conserving semi-numerical model SCRIPT}

\author[Maity \& Choudhury]{
Barun Maity$^{1}$\thanks{E-mail: bmaity@ncra.tifr.res.in}
and Tirthankar Roy Choudhury$^{1}$\\
$^{1}$National Centre for Radio Astrophysics, TIFR, Pune University Campus, Post Bag 3, Pune 411 007, India}
\date{Accepted XXX. Received YYY; in original form ZZZ}

\pubyear{2023}

\begin{document}
\label{firstpage}
\pagerange{\pageref{firstpage}--\pageref{lastpage}}
\maketitle

\begin{abstract}
One of the most promising probes to constrain the reionization history of the universe is the power spectrum of neutral hydrogen 21~cm emission fluctuations. The corresponding analyses require computationally efficient modelling of reionization, usually achieved through semi-numerical simulations. We investigate the capability of one such semi-numerical code, SCRIPT, to constrain the reionization parameters. Our study involves creating a mock data set corresponding to the upcoming SKA-Low, followed by a Bayesian inference method to constrain the model parameters. In particular, we explore in detail whether the inferred parameters are unbiased with respect to the inputs used for the mock, and also if the inferences are insensitive to the resolution of the simulation. We find that the model is successful on both fronts. We also develop a simple template model of reionization which can mimic the complex physical processes like inhomogeneous recombinations and radiative feedback and show that it can recover the global reionization history reliably with moderate computational cost. However, such simple models are not suitable for constraining the properties of the ionizing sources. Our results are relevant for constraining reionization using high-quality data expected in future telescopes. 
\end{abstract}




\section{Introduction}
\label{sec:introduction}

The epoch of reionization is that phase of the history of our universe that allows us to connect the early universe described by (almost) linear perturbations and the late stages, which are dominated by complex and non-linear structure formation and astrophysical processes. This is the epoch when the universe transits from a predominantly neutral to the mostly ionized state via the ionizing photons coming from the very first luminous sources \citep[for reviews, see][]{2001PhR...349..125B,2009CSci...97..841C,2018PhR...780....1D,2022arXiv220802260G,2022GReGr..54..102C}. Neither the exact timeline of the process of reionization is understood, nor do we have a good understanding of the nature of the first sources that reionized the universe.

The baryonic component of the universe consists mainly of hydrogen atoms. So, the redshifted 21~cm signal originating from the spin flip transition at the ground state of the neutral hydrogen atom is one of the most promising probes of reionization. In particular, the radio interferometric observations that are able to track the fluctuations in neutral hydrogen field should provide detailed information about the state of the intergalactic medium (IGM) during reionization. As the 21~cm signal is very faint, it is very hard to detect it with the sensitivities of the present observatories. Nevertheless, considerable progress has been achieved with the current interferometric observations, which have started to provide upper limits on the power spectrum of the cosmological 21~cm signal. These interferometers include Low Frequency Array \citep[LOFAR;][]{2019MNRAS.488.4271G,2020MNRAS.493.1662M}, Murchison Widefield Array \citep[MWA;][]{2019ApJ...884....1B,2020MNRAS.493.4711T}, Giant Metrewave Radio Telescope \citep[GMRT;][]{2013MNRAS.433..639P} and Precision Array for Probing the Epoch of Reionization \citep[PAPER;][]{2010AJ....139.1468P}. It is expected that the future telescopes like Square Kilometre Array \citep[SKA-Low;][]{2015aska.confE...1K} and Hydrogen Epoch of Reionization Array \citep[HERA;][]{2017PASP..129d5001D,2022ApJ...925..221A} will be able to pin down large scale power spectrum within $\sim 10\%$ uncertainties in about a few hundred hours of observations.

To correctly interpret these data, we need reliable modelling of the physics of reionization. There exist a variety of models, starting from the simple analytical ones to the very complex radiative simulations. The very early models of reionization would be rather simplistic, where the ionization field could be generated by approximating the ionized regions as spherical bubbles \citep{2004ApJ...613....1F,2005MNRAS.356.1519B,2007MNRAS.382..809D,2008ApJ...681..756S}. These models were useful to provide initial insights on the overall characteristics of the 21~cm power spectrum. On the other extreme in terms of complexities and computational requirements, full radiation hydrodynamic simulations have been developed by several groups \citep{2006MNRAS.372..679M,2006MNRAS.369.1625I,2007ApJ...671....1T,2015MNRAS.447.1806G,2016MNRAS.463.1462O,2019MNRAS.483.1029K,2020MNRAS.496.4087O,2022MNRAS.511.4005K, 2022MNRAS.512.4909G,2022arXiv220713098P} which capture the detailed physics of individual sources and surrounding regions. Although these simulations are useful for capturing the interplay between different physical processes during reionization, one major bottleneck is that they cannot be used to do parameter space exploration because of computational constraints.

The trade-off between accuracy and efficiency can be achieved by the semi numerical approaches. Instead of modelling the full radiative transfer, these models employ some kind of algorithm to count the photons and compare them with the hydrogen distribution to generate the ionization maps \citep{2007ApJ...669..663M,2008MNRAS.386.1683G,2010MNRAS.406.2421S,2011MNRAS.411..955M,2013ApJ...776...81B}.  A major advantage of these models is that they can be used to do parameter inference studies using the 21~cm power spectrum as one of the observational probes. In the absence of a real detection of the reionization power spectrum, one usually generates mock data expected from the next generation telescopes and studies the recovery of the model parameters along with the forecast for the errors on them using Bayesian inference techniques \citep{2015MNRAS.449.4246G,2017MNRAS.468..122H,2017MNRAS.472.2651G,2018MNRAS.477.3217G,2019MNRAS.484..933P,2021MNRAS.501.4748Q}. These calculations require the model to be evaluated numerous times as one samples the parameter space, thus the computational efficiency becomes extremely crucial in such studies. One can bypass the large number of evaluations through machine learning techniques, which also have been employed in forecasting the parameter constraints using the future 21~cm power spectra data \citep{2017ApJ...848...23K,2017MNRAS.468.3869S,2018MNRAS.475.1213S,2019MNRAS.490..371D,2022ApJ...933..236Z,2022MNRAS.512.5010C}. With the recent upper limits on the power spectrum measurements from different telescopes, these semi-numerical models have been exploited extensively to rule out somewhat extreme models of heating and ionization \citep{2015ApJ...809...62P,2016MNRAS.455.4295G,2020MNRAS.493.4728G,2020MNRAS.498.4178M,2021MNRAS.500.5322G,2021MNRAS.501....1G,2021MNRAS.503.4551G,2022ApJ...924...51A}.

Several of the above models are based on the excursion set method, perhaps the most popular and efficient method of implementing a semi-numerical photon counting algorithm \citep{2007ApJ...669..663M,2008MNRAS.386.1683G,2009MNRAS.394..960C,2011MNRAS.411..955M}. It has been known that the excursion set based models encounter the issue of photon number non-conservation, i.e., the number of ionizing photons produced by the sources is not equal to the number of hydrogen atoms ionized \citep{2007ApJ...654...12Z, 2011MNRAS.414..727Z,2016MNRAS.460.1801P}. More recently, \citet{2018MNRAS.481.3821C} have shown that, as a consequence of this photon non-conservation, the large scale 21~cm power spectra depend on the resolution of the simulation box used to generate the ionization fields. This non-convergence of the power spectrum has important consequences for inferring the properties of the IGM from the observational data. In general, one would feel inclined to use coarse resolution simulations while exploring the parameter space, as they require less computational resources. However, if the constraints are found to be dependent on the resolution of the simulations, then one needs to exercise extreme caution in setting up the simulation so that the inferred parameter values are not biased.

A possible solution to the photon non-conservation has been proposed via an explicitly photon conserving algorithm, namely \texttt{SCRIPT} \citep{2018MNRAS.481.3821C}. Although it has been shown that the model naturally produces numerically convergent power spectra at large scales, it still needs to be tested whether the recovered parameters remain unbiased when the resolution is varied. This will be the main aim of the work. Once this is confirmed, we also test if the model is able to efficiently constrain the reionization history when complex processes like the inhomogeneous recombinations and radiative feedback are included. Including these processes makes the code relatively inefficient, hence one needs to devise faster ways to explore the parameter space. To this end, we also develop an approximate template model which can mimic the complex effects through unknown parameters and still provide unbiased estimates of the global ionization fraction. As we will show in the paper, this provides a potential way to constrain the reionization history with relatively moderate computational resources.

The paper is organized as follows: In \secn{sec:mock_stat}, we provide a discussion on the generation of mock power spectra, followed by the formalism for the statistical analysis and parameter space exploration. We then introduce our most basic model of reionization in \secn{sec:two_param_model} and discuss the recovery of the corresponding model parameters. A more complex model of reionization is discussed in the subsequent \secn{sec:recom_feed}, where we also introduce a simple template reionization model for efficiently sampling the parameter space. Lastly, we summarize our results and discuss the consequences in \secn{sec:conc}. In this paper, the assumed cosmological parameters are $\Omega_m$ = 0.308, $\Omega_{\Lambda}$ = 0.691, $\Omega_b$ = 0.0482, $h$ = 0.678, $\sigma_8$ = 0.829 and $n_s$ = 0.961 \citep{2016A&A...594A..13P}.

\section{Mock data and statistical analysis}
\label{sec:mock_stat}

Let us first discuss our method for constructing the mock 21~cm data as appropriate for the upcoming experiments, and the procedure for constraining the model parameters by comparing with the data.

\subsection{The mock 21~cm power spectrum and telescope noise}
\label{sec:mock_powspec}

In the absence of actual data which can be analysed to constrain the reionization model parameters, one usually uses the theoretical models to construct the mock 21~cm data. The details of the model used to produce the 21~cm power spectrum will be discussed in the latter sections. In general, any model of reionization would produce the ionized hydrogen fraction $x_{\mathrm{HII}, i}$ in grid cells (represented by the index $i$) inside a simulation volume. The differential brightness temperature (assuming spin temperature is very much larger than CMB temperature) is given by \citep{1997ApJ...475..429M,2003ApJ...596....1C}
\be
\label{eq:delta_Tb}
\delta T_{b, i} \approx 27~\mathrm{mK}  \left(1 - x_{\mathrm{HII}, i}\right) \Delta_i \left(\frac{1+z}{10}\frac{0.15}{\Omega_{m}h^2}\right)^{1/2} \left(\frac{\Omega_{b}h^2}{0.023}\right),
\ee
where $\Delta_i \equiv \rho_{m, i} / \bar{\rho}_m$ is the ratio of the matter density $\rho_{m,i}$ in the grid cell and the mean matter density $\bar{\rho}_m$.

The observable we focus on in this paper is the dimensionless 21~cm power spectrum, defined as
\begin{equation}
\label{eq:Delta_21}
   \Delta_{21}^2(k) = \frac{k^3 P_{21}(k)}{2 \pi^2},
\end{equation}
where $P_{21}(k)$ is the power spectrum of the mean-subtracted fluctuation field $\delta T_{b,i} - \langle \delta T_{b, i} \rangle$.

The dominant contribution to the errors in the 21~cm power spectra comes from the thermal noise of the telescopes. In addition, we also need to account for the cosmic variance arising from surveying only a finite volume of the sky. These uncertainties on the mock 21~cm power spectra are obtained from a modified version of the publicly available package \texttt{21cmSense} \citep{2013AJ....145...65P,2014ApJ...782...66P}\footnote{The modified version of \texttt{21cmSense} used in this work can be found at \url{https://github.com/palc001/21cmSense}. This version has several new functions and also modules and data files to generate sensitivities for various telescopes.}. The mathematical details behind the computation of the interferometer sensitivities can be found in \citet{2012ApJ...753...81P}. It can be shown that the dimensionless power spectrum of the thermal noise is given by 
\be
\Delta_N^2(k) \approx X^2 Y\frac{k^3}{2\pi^2}\frac{\Omega}{2t}T_{\mathrm{sys}}^2,
\ee
where $X^2Y$ is the converting factor from bandwidths (frequency) and solid angles to cosmological comoving distances, $\Omega$ is the primary field of view, $t$ is the total integration time for Fourier mode $k$ and $T_{\mathrm{sys}}$ is the system temperature. After adding the cosmic variance term, the total noise in a bin $k_{\alpha}$ can be written as 
\be 
\label{eq:tot_noise}
\delta\Delta_T^2(k_{\alpha}) = \left(\sum_i \frac{1}{[\Delta_N^2(k_i)+\Delta_{\mathrm{21}}^2(k_i)]^2}\right)^{-\frac{1}{2}},
\ee
where $\Delta_{\mathrm{21}}^2(k)$ is the theoretical power spectrum as defined in \eqn{eq:Delta_21}, and the sum is over all independent Fourier modes $i$ that contribute to the particular bin labelled by $\alpha$.

The next step in creating the mock data is to add the noise and other uncertainties. This requires us to pick a telescope whose specifications would determine the noise. We choose SKA-Low specifications for this study \citep{2019arXiv191212699B}. The results we obtain would hold qualitatively for other upcoming telescopes, e.g. HERA, as well. For the telescope noise, we take 512 SKA-low stations with sizes of 40 metres in diameter\footnote{The antenna coordinates are taken from the SKA document summarizing the specifications, found at \url{https://www.skao.int/sites/default/files/documents/d18-SKA-TEL-SKO-0000422_02_SKA1_LowConfigurationCoordinates-1.pdf}}. The system temperature ($T_{\mathrm{sys}}$) is assumed to be $180~\mathrm{K} \left(\nu/180~\mathrm{MHz}\right)^{-2.5}$, where $\nu$ is the central frequency of observation in $\mathrm{MHz}$ units. We compute the noise for observations in a drift scan mode of 6 hours/day for 180 days, which gives a total observing time of around 1080 hours. The bandwidth is taken to be 8~MHz. We also assume a moderate foreground removal \citep{2014ApJ...782...66P} where the ``foreground wedge'' is considered to extend up to wave numbers $\Delta k_{\parallel} = 0.1 h~\mathrm{cMpc}^{-1}$ beyond the horizon limit, $k_{\parallel}$ being the component of the Fourier mode vector $\mathbf{k}$ along the line of sight. Note that the specifications chosen lead to an extremely low noise for the 21~cm power spectra and correspond to the best quality data we expect in the coming decade or so. Since the aim of our work is to understand the possible biases arising in the recovered parameters because of the assumptions made while modelling, we consider the most optimistic case in terms of error-bars in the model parameters and see if the recovery is within these error-bars. In this sense, this can be considered as possibly the most stringent test for the models.

In this work, we generate the mock data at three redshifts, $z=6.2$, $7$ and $8$ which correspond to observational frequencies $\approx 197$, $178$ and $158$~MHz, respectively. These redshifts allow us to probe different representative phases of the reionization history. We use power spectra in $k$-bins ranging from $0.102 h~\mathrm{cMpc}^{-1}$ to $1.047 h~\mathrm{cMpc}^{-1}$, the bins being linearly spaced at intervals of $\Delta k = 0.035 h~\mathrm{cMpc}^{-1}$. The $k$-range considered here are well suited for studying the cosmological signal from reionization, $k$-modes smaller than what we consider are dominated by the cosmic variance while the larger ones are dominated by the thermal noise.

To calculate the theoretical power spectrum, we use a simulation box of size $256h^{-1}\mathrm{cMpc}$ and generate the power spectra using a resolution of  $\Delta x = 2 h^{-1}\mathrm{cMpc}$. The box size is sufficiently large to ensure that the effects of missing Fourier modes are not substantial \citep{2014MNRAS.439..725I,2020MNRAS.495.2354K,2022arXiv220901225G} and it also covers the lowest $k$-bins we are interested in. The resolution of the simulation is set by a technical limitation of the way we generate the collapsed haloes in the box, see \secn{sec:two_param_model}. However, this does not cause any serious concern in our analysis because the higher $k$-modes are heavily dominated by the thermal noise and do not play any significant role in interpreting the cosmological signal.  The mock power spectrum used for further analysis is simply the theoretical power spectrum $\Delta_{21}^2(k_{\alpha})$ for a fiducial set of parameters plus a random number drawn from a Gaussian distribution having zero mean and standard deviation equal to the thermal noise $\Delta_N^2(k_{\alpha})$ for the particular $k$-bin. This shifting of the data points by adding appropriate random numbers is necessary because we do \emph{not} explicitly add the thermal noise to the generated 21~cm maps. In contrast, the scatter due to the cosmic variance is already included in the mock data when we carry out the parameter space exploration. This is because we use a realization of the initial density field for generating the mock power spectra different from the one used for computing the model 21~cm power spectra, see \secn{sec:two_param_model} for more details.

\subsection{Likelihood and parameter space exploration} 
\label{sec:parms_stat}

Given the mock data, we employ a Bayesian approach to recover the parameters of our model. The main goal is to compute the conditional probability distribution or the posterior $\mathcal{P}(\theta \vert \mathcal{D})$ of the model parameters $\theta$ given the mock data sets $\mathcal{D}$ mentioned in the previous section. This can be computed using the Bayes' theorem
\be\label{eq:bayes_eq}
 \mathcal{P}(\theta\vert \mathcal{D})=\frac{\mathcal{L}(\mathcal{D} \vert \theta) ~\pi(\theta)}{\mathcal{P}(D)},
\ee
where $\mathcal{L}(\mathcal{D} \vert \theta)$ is the conditional probability distribution of data given the parameters or the likelihood, $\pi(\theta)$ is the prior and $\mathcal{P}(\mathcal{D})$ is the evidence (which can be treated as the normalization parameter and does not play any role in our analysis). The likelihood is assumed to be multidimensional Gaussian which is similar to our earlier study \citep{2022MNRAS.515..617M}
\bear
\label{eq:chisq_eq}
\mathcal{L}(\mathcal{D} \vert \theta) 
&= \exp \left(-\frac{1}{2} \sum_{\alpha}\left[\frac{\Delta_{21,\mathcal{D}}^2(k_{\alpha})-\Delta_{21}^2(k_{\alpha}; \theta)}{\delta\Delta_T^2(k_{\alpha})}\right]^2 \right)
\nline
&= \prod_{\alpha} \exp \left(-\frac{1}{2} \left[\frac{\Delta_{21,\mathcal{D}}^2(k_{\alpha})-\Delta_{21}^2(k_{\alpha}; \theta)}{\delta\Delta_T^2(k_{\alpha})}\right]^2 \right),
\ear
where $\Delta_{21}^2(k_{\alpha}; \theta)$ are the model predictions for the parameters $\theta$, $\Delta_{21,\mathcal{D}}^2(k_{\alpha})$ are the mock 21~cm data points and $\delta \Delta_T^2(k_{\alpha})$ are the corresponding error bars on the data computed using \eqn{eq:tot_noise}. The summation index $\alpha$ runs over all $k$-values used in the analysis. 

We sample the posterior distribution using the  Markov Chain Monte Carlo (MCMC) method, more specifically, the Metropolis-Hastings algorithm \citep{1953JChPh..21.1087M}. We make use of the publicly available package \texttt{cobaya} \citep{2021JCAP...05..057T}\footnote{\url{https://cobaya.readthedocs.io/en/latest/}} to run the MCMC chains.  The samples are drawn using 20 parallel chains \citep{2002PhRvD..66j3511L,2013PhRvD..87j3529L}. The chains are assumed to converge when the Gelman-Rubin $R - 1$ statistic \citep{1992StaSc...7..457G} becomes less than a threshold $0.01$. This typically needs around $10^5$ steps for our model and completes in about 2-3 days for a resolution of $2 h^{-1}\mathrm{cMpc}$ with a $256 h^{-1}\mathrm{cMpc}$ box, and in $\sim$ 5-6 hours when the resolution is coarsened to $4 h^{-1}\mathrm{cMpc}$. We discard the first $30\%$ steps in the chains as `burn-in' and work with only the rest.

\section{Analysis with a two parameter reionization model}
\label{sec:two_param_model}

We now start the discussion on recovering the input model parameters by comparing with the mock data. In this work, we use the explicit photon conserving semi-numerical model \texttt{SCRIPT} \citep{2018MNRAS.481.3821C} to construct the mock data and for the likelihood analysis. In its simplest and most computationally efficient form, the reionization can be modelled at any given redshift using only two parameters. Generating the ionization field at a given redshift does not require any knowledge of the reionization history at earlier redshifts, hence the requirement of computational resources is minimal; we refer to such models as \textbf{single-snapshot} models. It is possible to extend \texttt{SCRIPT} to include several other physical effects, however that introduces several additional parameters, requires knowledge of the reionization and thermal histories and thus makes the model more computationally expensive. Models which require modelling the full history to compute the ionization maps at a given redshift will be referred to as \textbf{full-history} models.

In this section, we will describe our analysis using the simplest two parameter single-snapshot model and take up more complicated full-history cases in the next section. Let us first describe the main features of the model, for details we refer the reader to \citet{2018MNRAS.481.3821C}:

\begin{itemize}

\item \texttt{SCRIPT} provides the ionization field as the main output given two input fields at any redshift of interest. These input fields correspond to the dark matter density distribution and collapsed mass fraction in haloes which can produce ionizing photons. In general, these input fields can be got from any $N$-body dark matter only simulation. In case one is interested only at relatively large scale $\gtrsim 1 h^{-1}$~cMpc features of the IGM, it is sufficient to use 2LPT formalism to construct the density fields \citep{2011MNRAS.415.2101H}\footnote{\url{https://www-n.oca.eu/ohahn/MUSIC/}}. The collapsed fraction field is computed using subgrid prescription based on conditional ellipsoidal mass function \citep{2002MNRAS.329...61S}. The prescription requires mapping the non-linear density field (the Eulerian density) in the simulation volume to the initial linear density field (the Lagrangian density). We use the spherical approximation to carry out the mapping, however, the approximation breaks down for grid cells that are too small. We checked and found that the smallest grid cell we can use corresponds to a length $\Delta x = 2 h^{-1}$~cMpc, which sets the smallest scales we can probe.

\item In the most basic version of the code, there are only two free parameters which need to be specified. The first one is the ionizing photon efficiency, $\zeta$, which gives the available number of ionizing photons per unit number of hydrogen atoms. The other one is the minimum threshold mass $M_{\mathrm{min}}$ for haloes that can contribute to the ionizing photon budget. The number density of ionizing photons produced in a grid cell $i$ is given by
\be
n_{\mathrm{ion}, i} = \zeta f_{\mathrm{coll}, i}(M_{\mathrm{min}})~n_{H, i},
\ee
where $f_{\mathrm{coll}, i}(M_{\mathrm{min}})$ is the fraction of mass in collapsed haloes with masses $\geq M_{\mathrm{min}}$ and $n_{H, i}$ is the hydrogen number density in the cell. The above expression can be easily generalized to cases where $\zeta$ depends on the halo mass, see \citet{2022MNRAS.511.2239M} for such models.

\item The photon conserving algorithm consists of two steps. In the first, photons from grid cells containing ionization sources are distributed to neighbouring cells in increasing order of distance until all the photons are exhausted. This process is carried out independently for all source cells, and hence may lead to cells where the ionization fraction exceeds unity. In the second step, which actually consists of a series of substeps, these excess photons in overionized cells are redistributed to nearby cells. The process converges when there are no unphysical overionized cells left in the volume.

The ionization condition of a cell is determined by the number density of ionizing photons available $n_{\mathrm{ion, avail}, i}$ and $n_{H, i}$. The number of available ionizing photons can be written as
\be
\label{eq:nion_avail}
n_{\mathrm{ion, avail}, i} = \sum_j n_{\mathrm{ion}, j \to i},
\ee
where $n_{\mathrm{ion}, j \to i}$ is the number of photons contributed by the cell $j$ to cell $i$. Clearly, $n_{\mathrm{ion}, j \to i}$ will be dominated by cell pairs whose distance is small, and will also depend on the density distribution of cells situated between $i$ and $j$. At any step of the algorithm, a cell is assigned fully ionized if
\be
\label{eq:nion_geq_nH}
n_{\mathrm{ion, avail}, i} \geq n_{H, i},
\ee
and the excess photons distributed to other cells in the next step. Other cells are assigned an ionized fraction $x_{\mathrm{HII}, i} = n_{\mathrm{ion, avail}, i} / n_{H, i}$.

\item Once we obtain the ionization fraction $x_{\mathrm{HII},i}$ for each cell in the simulation box, the global mass averaged ionization fraction $Q^M_{\mathrm{HII}} = \langle x_{\mathrm{HII},i}\Delta_i\rangle$ can be derived, the angular brackets denoting the average over all grid cells in the box. The photon conserving criterion establishes that the total average number of ionizing photons per hydrogen atoms is equal to the global ionization fraction, i.e., $\langle \zeta f_{\mathrm{coll},i}(M_{\mathrm{min}})~\Delta_i\rangle = Q^M_{\mathrm{HII}}$. One important advantage of the photon conserving model is that it produces large scale 21~cm power spectra which are independent of resolutions of the simulation box, a feature we will study next.

\end{itemize}

\subsection{Recovery of parameters}

\begin{figure}
    \includegraphics[width=0.99\columnwidth]{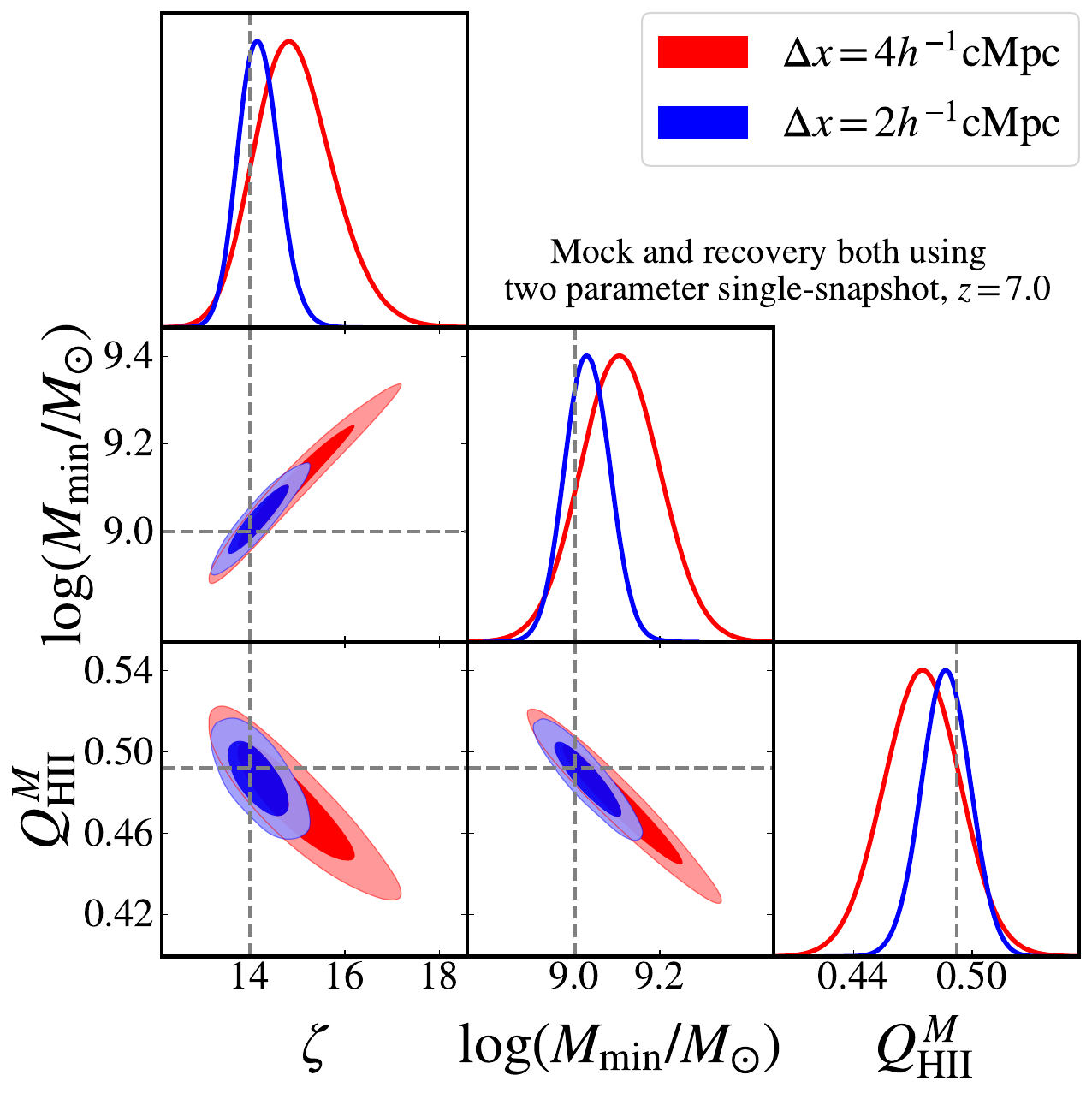}
    \caption{Recovery of parameters using the photon conserving two parameter single-snapshot model at redshift $z=7$. The mock data is created by the same single-snapshot model (but with a different realization of the matter density field) using a grid resolution $\Delta x = 2 h^{-1}$~cMpc. The off diagonal panels show the joint two-dimensional posterior distributions of each pair of parameters. The contours represent $68\%$ and $95\%$ confidence intervals. The diagonal panels show the marginalized posterior distributions of the parameters. The dashed lines represent the input values for generating mock data. The \textit{blue} contours and curves show the results when the resolution used for the MCMC run is $\Delta x = 2 h^{-1}$~cMpc (same as the one used for the mock data), while the \textit{red} ones are for the resolution $\Delta x = 4 h^{-1}$~cMpc. The recovery of the input parameters are unbiased irrespective of the resolution used.}
    \label{fig:two_params_7}. 
\end{figure}

\begin{figure}
    \includegraphics[width=0.5\textwidth]{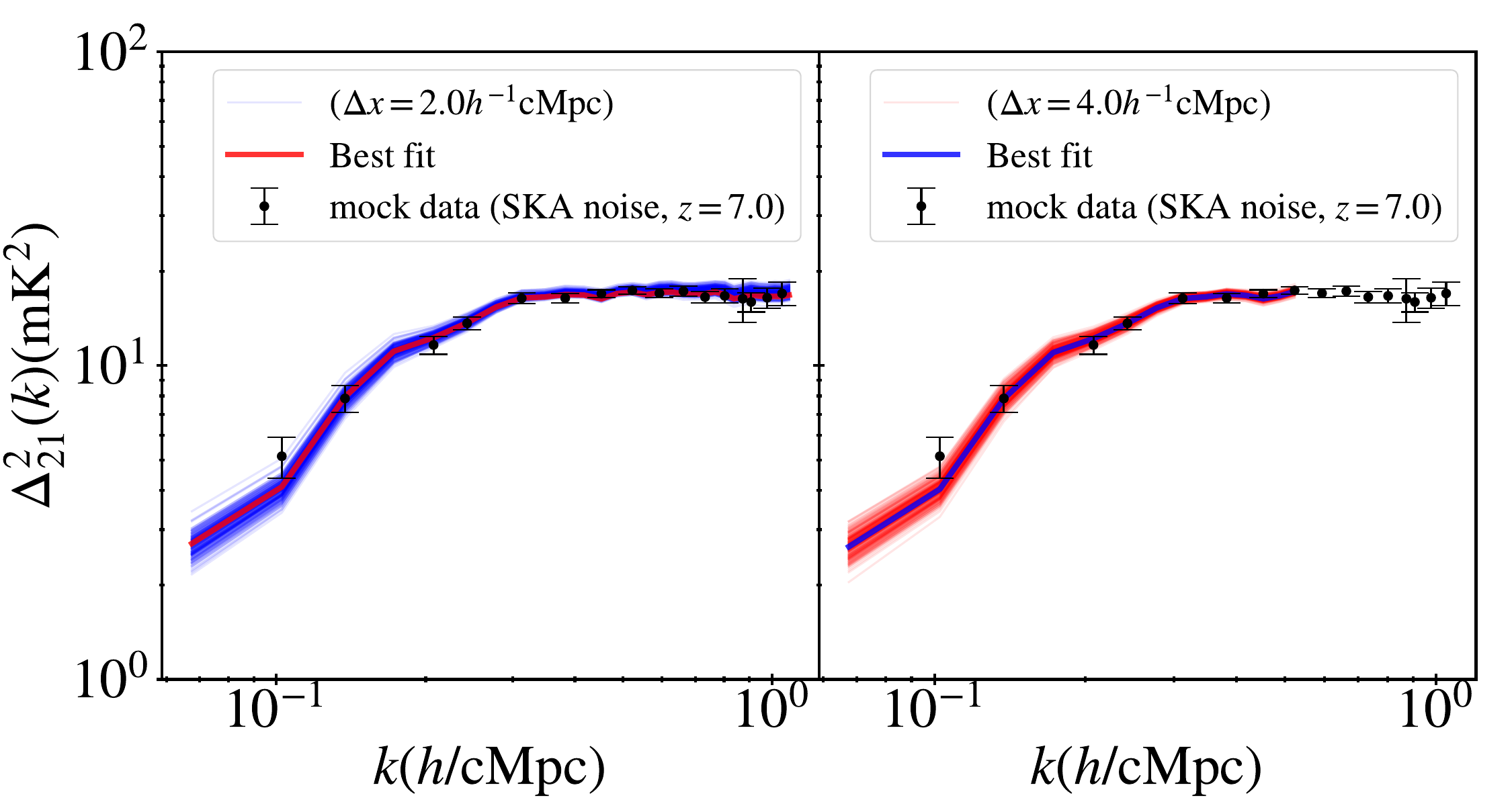}
    \caption{The 21~cm power spectra for the photon conserving two parameter single-snapshot model at redshift $z=7$ using 200 random samples drawn from the posterior distributions shown in \fig{fig:two_params_7}. The \textit{blue} lines show the models for a resolution $\Delta x=2 h^{-1} \mathrm{cMpc}$ while the \textit{red} ones are for $\Delta x=4h^{-1}\mathrm{cMpc}$. The dashed lines denote the input models.}
    \label{fig:two_params_pow_7}
\end{figure}

\begin{figure}
    \includegraphics[width=0.99\columnwidth]{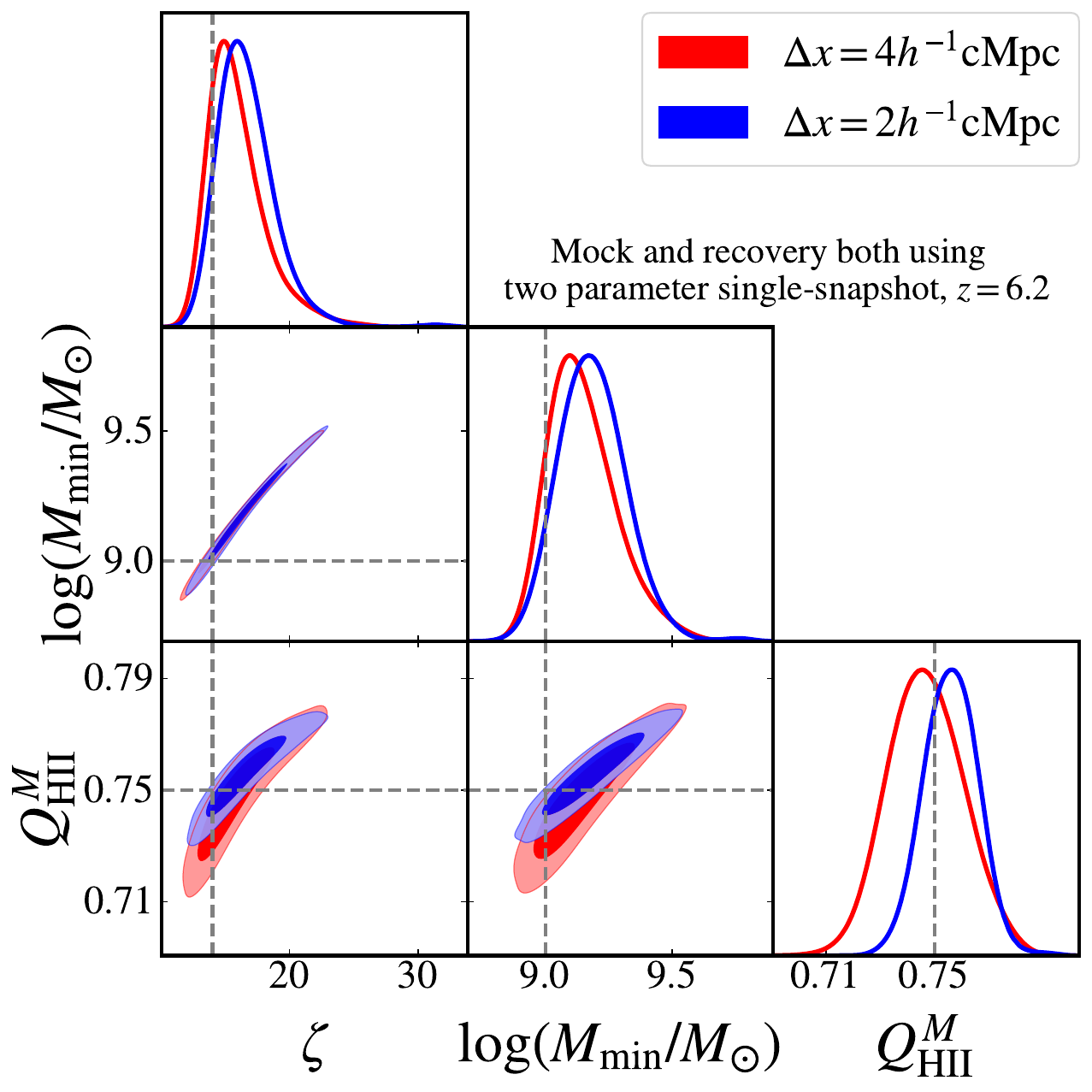}
    \caption{Same as \fig{fig:two_params_7} but for redshift $z=6.2$.}
    \label{fig:two_params_62}
\end{figure}

\begin{figure}
    \includegraphics[width=0.5\textwidth]{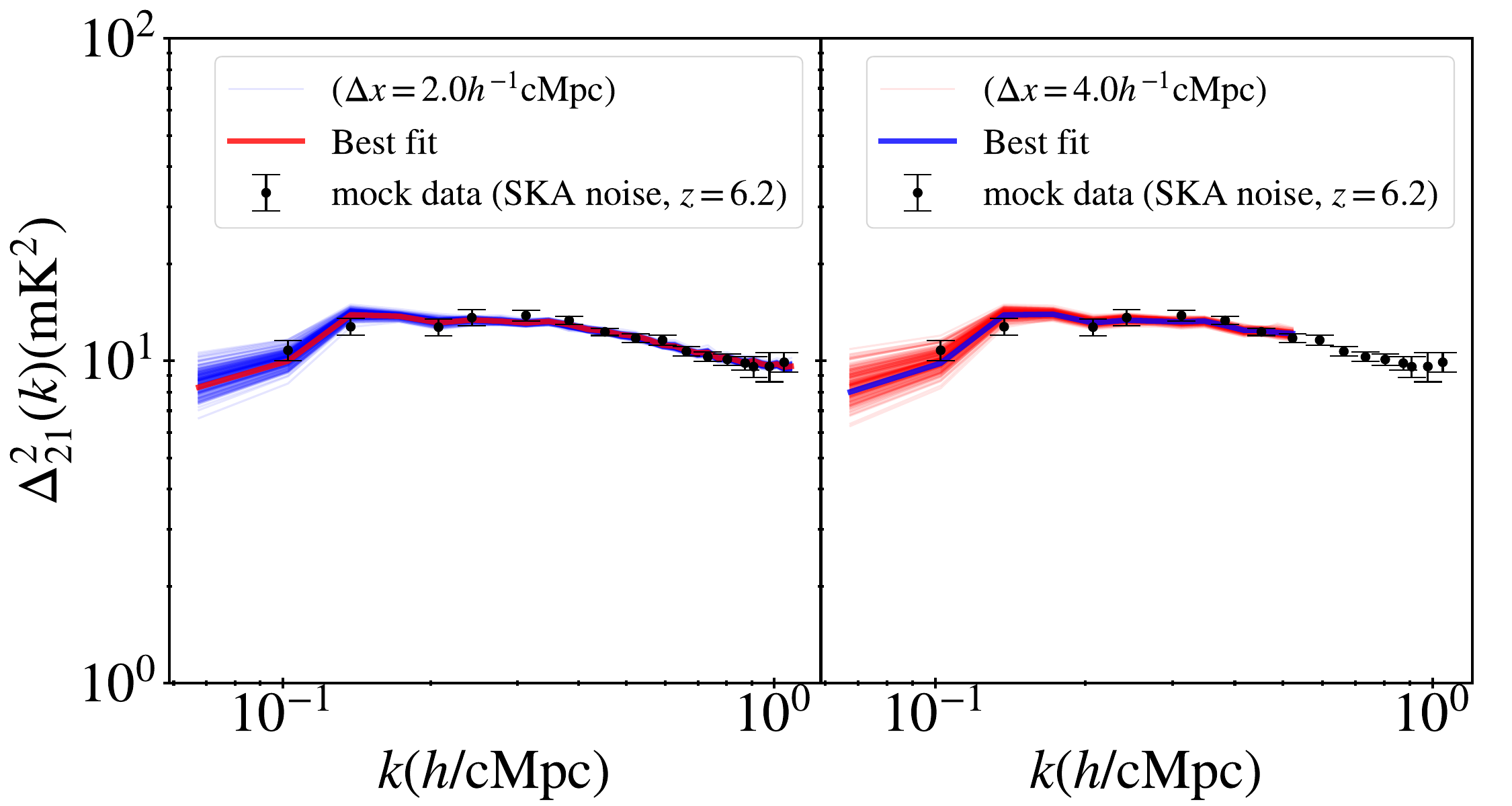}
    \caption{Same as \fig{fig:two_params_pow_7} but for redshift $z=6.2$.}
    \label{fig:two_params_pow_62}
\end{figure}

\begin{figure}
    \includegraphics[width=0.99\columnwidth]{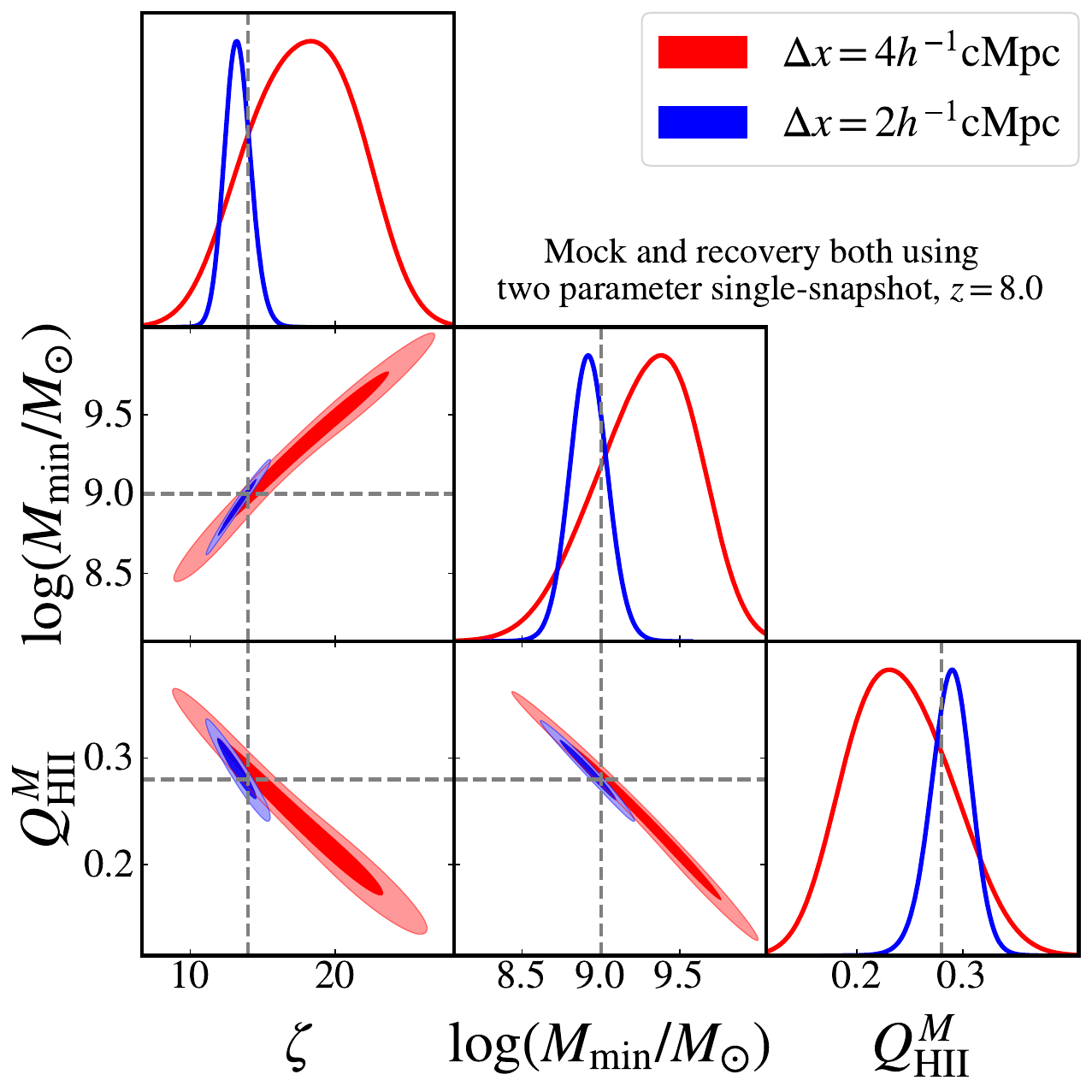}
    \caption{Same as \fig{fig:two_params_7} but for redshift $z=8$.}
    \label{fig:two_params_8}
\end{figure}

\begin{figure}
    \includegraphics[width=0.5\textwidth]{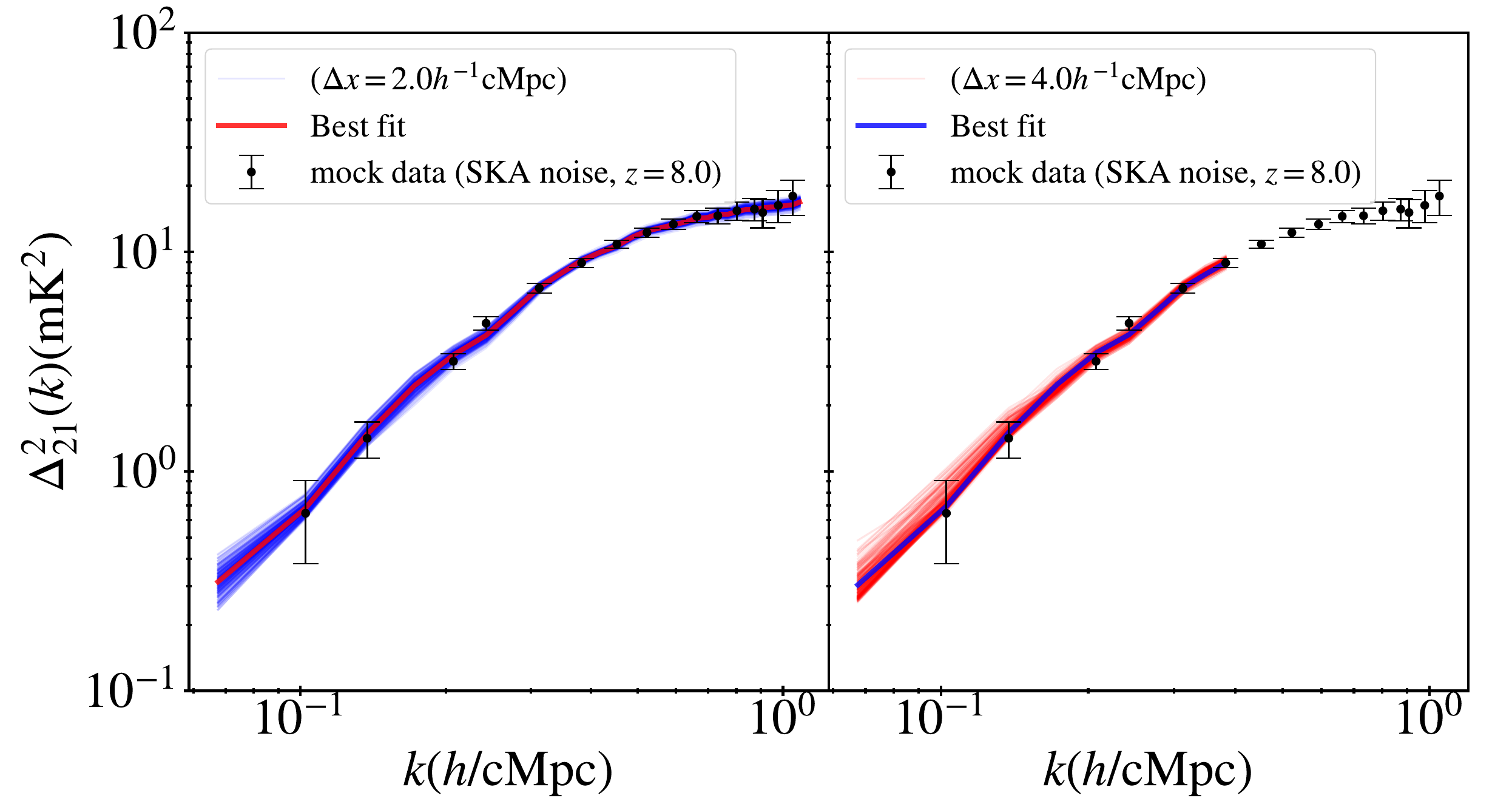}
    \caption{Same as \fig{fig:two_params_pow_7} but for redshift $z=8$.}
    \label{fig:two_params_pow_8}
\end{figure}

\begin{table*}
\renewcommand{\arraystretch}{1.4}
\begin{tabular}{|c|c|c|c|c|c|c|}
\hline
Parameter & Prior & \multicolumn{5}{c}{Photon conserving two parameter single-snapshot model}\\ %
  
\hline
\multicolumn{3}{|c|}{$z=7.0$}  & \multicolumn{2}{|c|}{$\Delta x=2h^{-1}\mathrm{cMpc}$} & \multicolumn{2}{|c|}{$\Delta x=4h^{-1}\mathrm{cMpc}$} \\
\cline{1-7}
& & input & mean [68$\%$ C.L.] & best-fit  & mean [68$\%$ C.L.] & best-fit \\

\hline
 $\zeta$ &[2, 100] &14 & $14.187^{+0.403}_{-0.439}$ & 14.173  & $14.955^{+0.726}_{-0.899}$ & 14.874  \\

 $\log(M_{\mathrm{min}} / M_{\odot})$ & [7, 11] & 9 & $9.029^{+0.052}_{-0.052}$ & 9.029  & $9.107^{+0.093}_{-0.092}$ & 9.103 \\
 $Q_{\mathrm{HII}}^M$ &[0.1,1] & 0.49 & $0.487^{+0.012}_{-0.012}$ & 0.487  & $0.475^{+0.019}_{-0.019}$ & 0.475 \\
 $\chi^2 / \nu$ &$-$ &$-$ & $-$ & $6.519/14$ & $-$   &  $4.175/6$ \\
\hline
\multicolumn{3}{|c|}{$z=8.0$}  & \multicolumn{2}{|c|}{$\Delta x=2h^{-1}\mathrm{cMpc}$} & \multicolumn{2}{|c|}{$\Delta x=4h^{-1}\mathrm{cMpc}$} \\
\hline
 $\zeta$ &[2, 100] &14 & $13.266^{+0.795}_{-0.917}$ & 13.111  &$17.824^{+3.955}_{-3.690}$ &15.120 \\

 $\log(M_{\mathrm{min}} / M_{\odot})$ & [7, 11] & 9 & $8.923^{+0.116}_{-0.112}$ & 8.905  & $9.289^{+0.362}_{-0.282}$ & 9.090 \\
 $Q_{\mathrm{HII}}^M$ & [0.1,1] & 0.28 & $0.289^{+0.019}_{-0.017}$ & 0.292  & $0.240^{+0.046}_{-0.053}$ & 0.273\\
 $\chi^2 / \nu$ &$-$ &$-$ & $-$ & $4.875/14$ & $-$   &$3.939/4$ \\
\hline
\multicolumn{3}{|c|}{$z=6.2$}  & \multicolumn{2}{|c|}{$\Delta x=2h^{-1}\mathrm{cMpc}$} & \multicolumn{2}{|c|}{$\Delta x=4h^{-1}\mathrm{cMpc}$} \\
\hline
 $\zeta$ &[2, 100] &14 & $16.641^{+1.604}_{-2.517}$ & 15.466 &$15.874^{+1.364}_{-2.579}$ & 14.907\\

 $\log(M_{\mathrm{min}} / M_{\odot})$ & [7, 11] & 9 & $9.186^{+0.120}_{-0.144}$ & 9.125 & $9.146^{+0.102}_{-0.157}$ & 9.093\\
 $Q_{\mathrm{HII}}^M$ &[0.1,1] & 0.75 & $0.756^{+0.010}_{-0.010}$ & 0.751  & $0.747^{+0.013}_{-0.015}$ & 0.743\\
 $\chi^2 / \nu$ &$-$ &$-$ & $-$ & 10.191/14 & $-$   &  $8.456/6$ \\
\hline 
\end{tabular}
\caption{Parameter constraints obtained using the photon conserving two parameter single-snapshot model at redshifts $z=7.0,~8.0$ and $6.2$. Results are shown for two different resolutions $\Delta x = 2 h^{-1}\mathrm{cMpc}$ and $4 h^{-1}\mathrm{cMpc}$ used for the MCMC analysis. The mock data is generated using the same single-snapshot model, always at the resolution $\Delta x = 2 h^{-1}\mathrm{cMpc}$. For each parameter, we show the prior and the input value used for the mock along with the obtained mean, $68\%$ confidence limits and best-fit. We also provide the $\chi^2 / \nu$ for the best-fit models, $\nu$ being the number of degrees of freedom. 
}
\label{tab:table_1}
\end{table*}

We now investigate the recovery of model parameters for the two parameter single-snapshot model. The mock 21~cm power spectra are calculated using fiducial values of the free parameters, namely, $\zeta = 14$ and $M_{\mathrm{min}} = 10^9 M_{\odot}$. These choices lead to ionization fractions $Q_{\mathrm{HII}}^M = 0.28, 0.49, 0.75$ at redshifts $z = 8.0, 7.0, 6.2$ respectively.

While sampling the parameter space via the MCMC analysis, we use the same two parameter single-snapshot model, but with an initial density field realization different from the one used to construct the mock data. Let us first take the case where the simulation volume has the same resolution $\Delta x = 2 h^{-1}\mathrm{cMpc}$ as the one used to generate the mock data. We study the redshift $z = 7$ first, which corresponds to a $\sim 50\%$ ionized IGM. In \fig{fig:two_params_7}, we show the recovered parameter posteriors for this case (blue contours and curves). We find that the input parameters ($\zeta = 14$ and $M_{\mathrm{min}} = 10^9 M_{\odot})$, shown by dashed lines in the figure, are well within the $1 \sigma$ confidence levels of the posterior distribution. We also derive the distribution of $Q_{\mathrm{HII}}^M$ and find that it peaks around the input value of $\approx 0.5$. The degeneracies between the parameters are straightforward to understand: $\zeta$ and $M_{\mathrm{min}}$ are positively correlated as higher values of $\zeta$ lead to larger number of ionizing photons per halo, and that can be compensated by increasing $M_{\mathrm{min}}$ so that the number of haloes decrease \citep{2015MNRAS.449.4246G,2022MNRAS.514L..31M}. The degeneracies between $Q_{\mathrm{HII}}^M$ and the two free parameters too can be understood from similar arguments. In the left-hand panel of \fig{fig:two_params_pow_7}, we show the plots of power spectra for 200 random samples from our MCMC chains (blue curves), along with the best-fit and the input models. The points with error-bars are the mock data. The match between the recovered power spectrum and the mock data is quite good. The values of the recovered parameters along with the statistical errors are listed in \tab{tab:table_1}. It is clear that the input values are always within the $1 \sigma$ of the recovered parameters. For completeness, we also provide the values of $\chi^2 / \nu$ for the best-fit model in the table, $\nu$ being the number of degrees of freedom. The values confirm that the match between the recovered best-fit model and the mock data is excellent.

Let us now check the recovery when the analysis is carried out with a simulation volume of coarser resolution $\Delta x = 4h^{-1}\mathrm{cMpc}$. Note that the mock data remains identical as before, i.e., generated using a finer resolution $\Delta x = 2 h^{-1}\mathrm{cMpc}$. This analysis requires less grid cells (for the same simulation volume) and hence is much more efficient than the previous one. The price one has to pay is that the data at $k \gtrsim 0.5 h / \mathrm{cMpc}$ cannot be used. This may not be a serious handicap as the mock data at most of the high-$k$ bins have relatively larger error-bars and hence would not contribute significantly to the likelihood.

As can be seen from \fig{fig:two_params_7} (red contours and curves), the recoveries for this coarser resolution are almost identical to those obtained using the finer resolution. The corresponding limits on the parameters can be found \tab{tab:table_1}, see the two right-most columns. The constraints for the coarser resolution case are slightly weaker than the finer one as we are not able to use the data points at very small-scales (large $k$ modes), e.g., the errors on $Q_{\mathrm{HII}}^M$ are $\sim 4\%$ for the coarser resolution compared to $\sim 2.5\%$ for the finer resolution.

This analysis confirms the numerical convergence of our photon conserving algorithm with respect to the grid size used. In particular, we confirm that using a coarser resolution does not lead to any bias in the inferred values of the parameters. This result is significant because it allows us to obtain unbiased constraints on the parameters using relatively coarser resolution and thus moderate computing requirements. In an earlier work, \citet{2018MNRAS.481.3821C} have shown that the non-convergence of the large-scale power spectrum with respect to the resolution is directly related to non-conservation of photons in semi-numerical models of reionization. Because the model used above is photon conserving by construction, the convergence is not surprising. However, for photon non-conserving models, e.g., those based on excursion set formalism, this convergence is not guaranteed. We study this in detail in \appndx{app:es_based} and quantify the bias in the recovered parameters for coarser resolution maps in excursion set-based models.

The next obvious step is to extend the analysis to other redshifts with different global ionization fractions. The posterior distributions and the corresponding sampled power spectra for a lower redshift ($z=6.2$) with a higher ionization fraction ($Q_{\mathrm{HII}}^M \approx 0.75$) are shown in \figs{fig:two_params_62} and \ref{fig:two_params_pow_62} respectively, the corresponding parameter values and statistical errors are quoted in \tab{tab:table_1}. As seen from the plots, the recovery is quite good, and that too for both the resolutions, thus confirming the generic nature of the numerical convergence of our photon conserving algorithm. Lastly, we show  the results for a higher redshift ($z=8.0$), i.e., a lower ionization state of the universe ($Q_{\mathrm{HII}}^M \approx 0.28$) in \figs{fig:two_params_8} and \ref{fig:two_params_pow_8} respectively, see \tab{tab:table_1} for the parameter values and errors. As with the other cases, the recovered parameters match the corresponding input values within the $1 \sigma$ for almost all the parameters. The uncertainties on the parameters are relatively larger when the resolution is coarser. This is expected as only the large-scale models are available for the likelihood calculations, and the power spectrum is rather featureless at these scales.

\section{Model with recombination and feedback}
\label{sec:recom_feed}
The model presented so far is very basic in nature, which does not include several physical processes related to reionization. In this section, we check the possibility of recovering the reionization parameters when the model consists of more complex processes.

\subsection{The full-history model}
\label{sec:full_history}

As is well known, reionization is associated with other inhomogeneous astrophysical processes like the recombination of free electrons and ionized atoms, and the radiative feedback on star forming haloes due to heating of the medium. Recently, we have extended our model to include these inhomogeneous effects, see \citet{2022MNRAS.511.2239M,2022MNRAS.515..617M}. The details of the model can be found in these two papers, we summarize the main features of the model here:

\begin{itemize}

\item The temperature of each grid cell in the simulation volume is computed using the appropriate evolution equation. We account for processes like the expansion cooling, adiabatic heating/cooling from evolution of the densities, photoheating and Compton cooling. The calculation requires introduction of a free parameter, namely, the reionization temperature increment $T_{\mathrm{re}}$. This parameter quantifies the increase in temperature when a region transitions from being fully neutral to a fully ionized state, i.e., it is the temperature of the region immediately after the ionization. The calculation of the photoheating rate requires knowledge of $T_{\mathrm{re}}$.

\item The number density of recombinations $n_{\mathrm{rec}, i}$ in each cell is computed self-consistently by tracking the ionization history of that cell. We account for the subgrid clumping of the IGM by introducing another free parameter $C_{\mathrm{HII}}$, which is the globally averaged clumping factor. In the presence of recombinations, the condition for assigning a cell to be ionized, \eqn{eq:nion_geq_nH}, is modified to
\be
n_{\mathrm{ion, avail}, i} \geq n_{H, i} + n_{\mathrm{rec}, i},
\ee
with the rest of the algorithm for generating ionization maps remaining unchanged.

\item We implement radiative feedback suppressing star formation in low mass haloes using a Jeans mass-based prescription. The prescription relies on the value of the temperature at each grid cell, and hence the amount of feedback too varies from cell to cell. So, the minimum halo mass for a cell (say $i$) is given by $M_{\mathrm{min}, i} = \mathrm{Max} \left[M_{\mathrm{cool}}, M_{J,i}\right]$, where $M_{\mathrm{cool}}$ is the minimum mass of haloes that can cool via atomic transitions and $M_{J,i}$ is the Jeans mass at virial overdensity. The Jeans mass depends upon the temperature of the region ($\propto T^{3/2}$) and is higher than $M_{\mathrm{cool}}$ in the ionized regions. It is thus clear that the $M_{\mathrm{min}}$ is determined by the atomic cooling in the neutral regions and by Jeans mass in the feedback affected ionized regions. The ionizing photon production rate of a cell then can be computed by summing over weighted contribution from neutral and ionized regions within a cell. Note that the minimum mass of haloes that can produce ionizing photons is not a free parameter any more, it is given by the atomic cooling condition in neutral regions and by the radiative feedback in the ionized regions.

\item It must be emphasized that the ionization map at a given redshift depends on the thermal and ionization history of the IGM. This makes the model very different from its basic form of \secn{sec:two_param_model}. While the maps for the two parameter single-snapshot model at a given redshift can be generated without any knowledge of the history of that cell, that is not possible for this extended full-history model any more. Consequently, the model becomes much slower computationally.

\end{itemize}

Upon comparing the theoretical model with measurements of the CMB optical depth \citep{2020A&A...641A...6P}, the dark pixel fraction at $z \sim 6$ \citep{2015MNRAS.447..499M}, UV luminosity function of $z \sim 6$ and $7$ galaxies \citep{2015ApJ...803...34B,2017ApJ...843..129B} and the low-density IGM temperature measurements \citep{2020MNRAS.494.5091G}, we find that a model with $\zeta(z) = 8 [10 / (1 + z)]^{2.3}$, $T_{\mathrm{re}} = 2 \times 10^4$~K and $C_{\mathrm{HII}} = 3$ is close to the model that best fits the data. The CMB optical depth for this model turns out to be $\tau_e = 0.053$. These fiducial values are used to construct the mock power spectra data for further analysis in this section.

\subsection{Recovery of parameters using the two parameter single-snapshot model}

\begin{figure}
    \includegraphics[width=0.91\columnwidth]{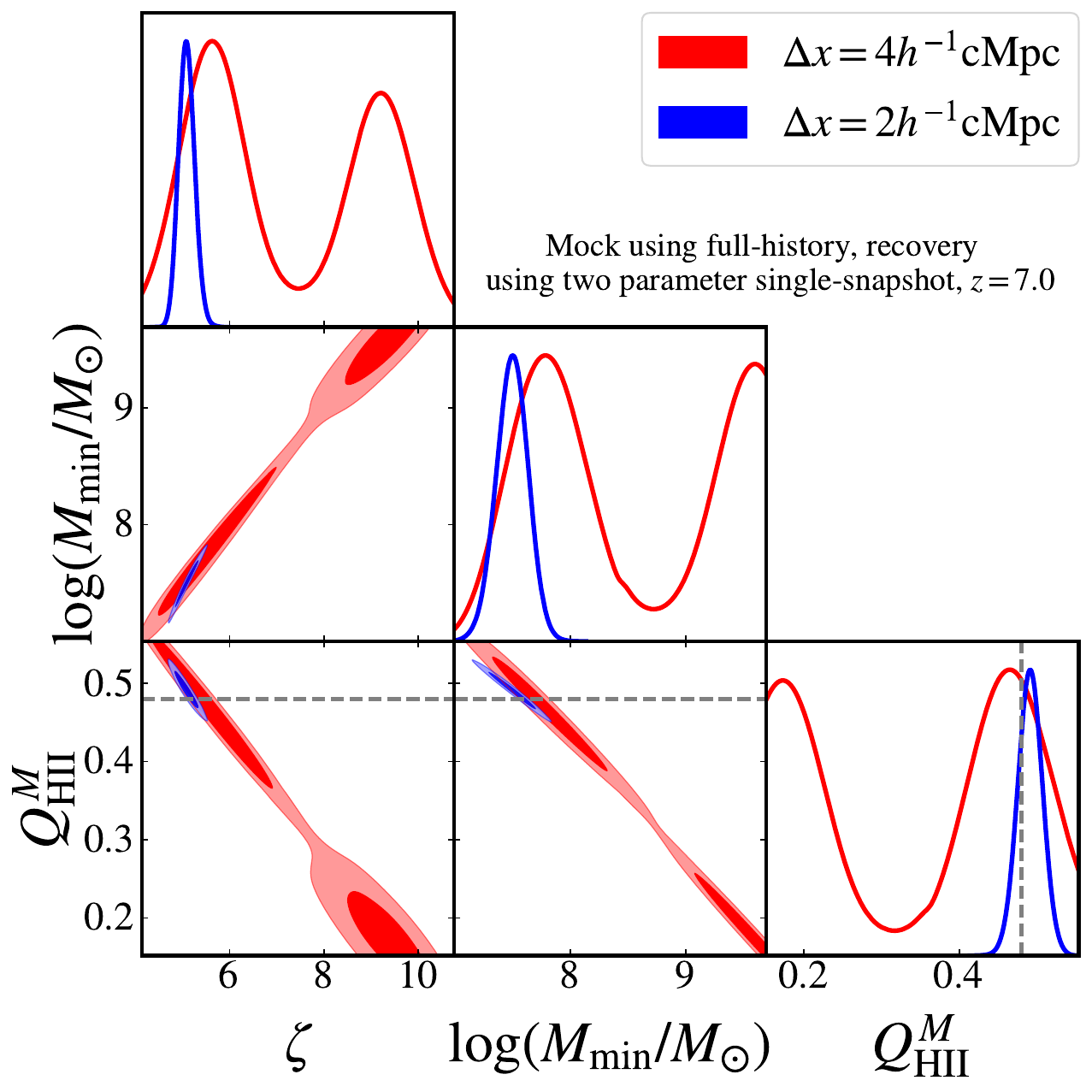}
    \caption{Recovery of parameters for the case where the mock 21~cm power spectrum data is generated using the full-history model (consisting of radiative feedback and inhomogeneous recombinations), while the one used for the parameter space exploration is the simple two parameter single-snapshot model. The descriptions of the different panels are the same as in \fig{fig:two_params_7}. Although the recovery of the input ionized fraction $Q_{\mathrm{HII}}^M$ is unbiased when the resolution used for parameter constraints is equal to that used for the mock, the recovery fails for the coarser resolution (red contours and curves).}
    \label{fig:two_params_feed_rec_z_7}
\end{figure}

Since the mock data has been generated using a rather complex (albeit semi-numerical) model of reionization, it is natural to study the recovery of the parameters using the same model. However, since computing the 21~cm power spectrum at a given redshift requires computing the full history, the model is not computationally efficient, and the parameter space exploration can be quite demanding. We thus take a different approach and try to recover the essential parameters of reionization using a template model which can be used at a given redshift snapshot without solving for the full history. A full parameter space exploration using the detailed model would require some sophisticated interpolation scheme in the parameter space (e.g., using emulators), which we postpone for a future work.

The first template model we try is our simple two parameter model of \secn{sec:two_param_model}. Clearly, this is a very simplistic reionization model where we have neglected the inhomogeneities in the recombinations and the effect of radiative feedback on the small mass haloes. In particular, the two parameters of the model $\zeta$ and $M_{\mathrm{min}}$ do not have any obvious counterparts in the detailed model. Hence, the constraints obtained on these parameters cannot be compared with any ``input'' values. What we can check is that whether this two parameter model can recover the value of $Q_{\mathrm{HII}}^M$ as implied by power spectra generated using the detailed model. 

We start with redshift $z=7.0$ where the global ionization fraction $Q_{\mathrm{HII}}^M \approx 0.5$ for the input model. In \fig{fig:two_params_feed_rec_z_7}, the blue regions and curves show the parameter recoveries using the resolution $\Delta x = 2h^{-1}\mathrm{cMpc}$, same as what was used to generate the mock data. As is clear from the figure, the recovered posterior distribution of $Q_{\mathrm{HII}}^M$ matches the input $Q_{\mathrm{HII}}^M$ surprisingly well. This indicates that the 21~cm power spectrum can be used for recovering the global ionization fraction even when the physical model uses simplistic descriptions of the underlying physical processes.

We next study the parameter recovery for the coarser resolution $\Delta x = 4 h^{-1}\mathrm{cMpc}$, see the regions and curves in red in \fig{fig:two_params_feed_rec_z_7}. Interestingly, we find a bimodality in the posterior distributions of the parameters for this case. In particular, the posterior of $Q_{\mathrm{HII}}^M$, our main quantity of interest, cannot be constrained reliably because of the bimodality. One of the peaks of the bimodal distribution is around the input value, however, the other peak is at a much lower value $Q_{\mathrm{HII}}^M \approx 0.2$. The strengths of both the peaks are almost similar, as we have checked from the values of the $\chi^2$. The reason for this bimodality is as follows: since the coarse resolution maps probe only the larger scales and also since the 21~cm power spectrum at large scales is non-monotonic \citep[see Figure 10 of][]{2022MNRAS.511.2239M}, the model ends us producing very similar power spectra at two widely different stages of reionization. 

We thus conclude that our two parameter single-snapshot model cannot be used reliably to describe the full-history model at different resolutions, as it cannot recover the value of $Q_{\mathrm{HII}}^M$ corresponding to the input model. Hence, we do not explore this model any further and move on to a slightly complicated template model which can possibly describe the detailed full-history model.

\subsection{A four parameter single-snapshot template}
\label{sec:toy_fb}

\begin{figure*}
    \includegraphics[width=0.75\textwidth]{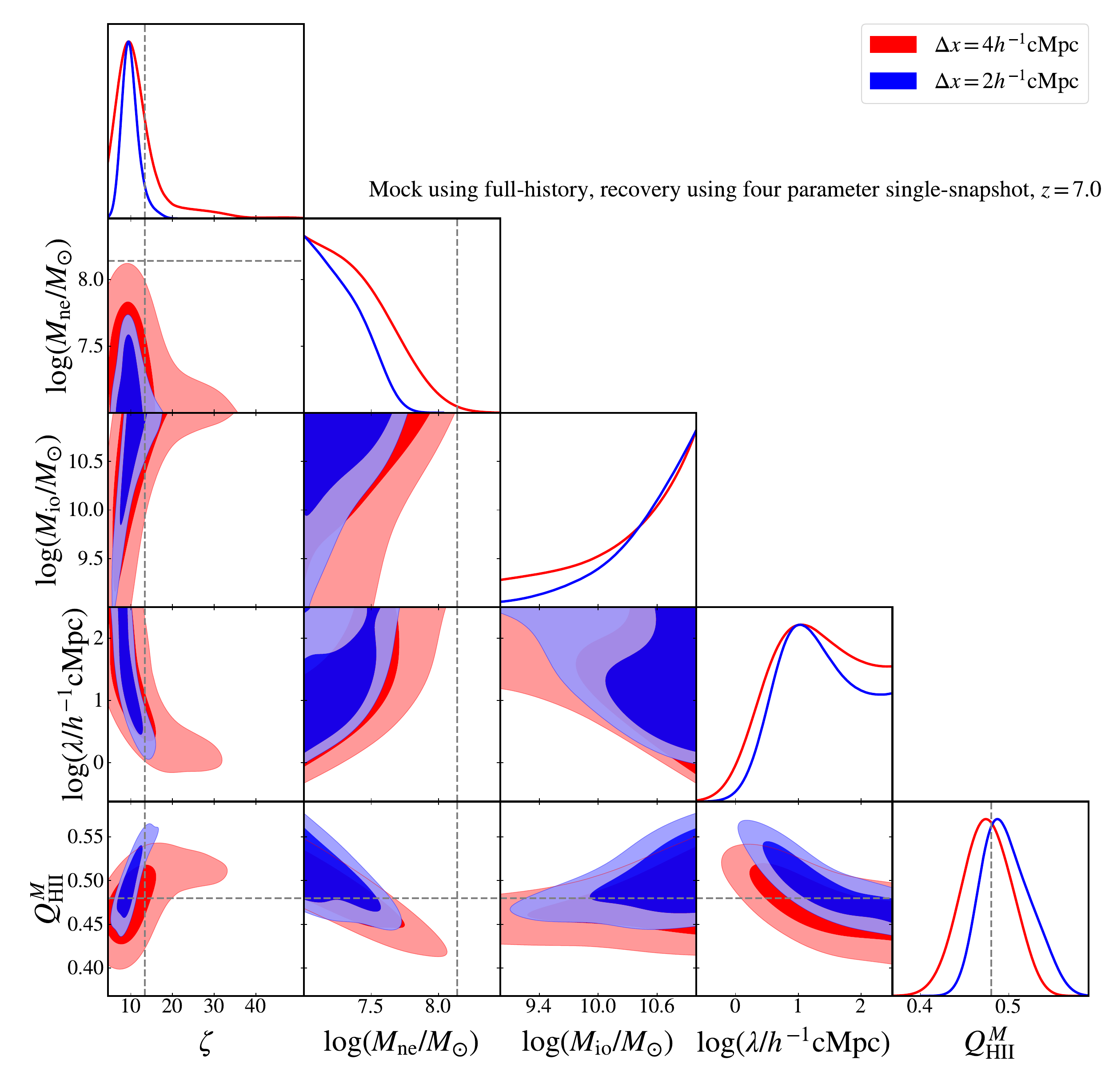}
    \caption{Recovery of parameters for the case where the mock 21~cm power spectrum data is generated using the full-history model (consisting of radiative feedback and inhomogeneous recombinations), while the one used for the parameter space exploration is the four parameter single-snapshot model. The off diagonal panels show the joint two-dimensional posterior distributions of each pair of parameters. The contours represent $68\%$ and $95\%$ confidence intervals. The diagonal panels show the marginalized posterior distributions of the parameters. Wherever possible, we show the input values for generating mock data by dashed lines. The \textit{blue} contours and curves show the results when the resolution used for the MCMC run is $\Delta x = 2 h^{-1}$~cMpc (same as the one used for the mock data), while the \textit{red} ones are for the resolution $\Delta x = 4 h^{-1}$~cMpc. The recovery of the parameters describing the reionization sources are biased with respect to the input, however, the global ionized fraction $Q_{\mathrm{HII}}^M$ is faithfully recovered for both the resolutions.}
    \label{fig:toy_recom_feed_z7}
\end{figure*}

It is obvious that the two parameter single-snapshot model cannot recover the global ionization fraction reliably because it is not informed of the effects of  inhomogeneous recombination and feedback. To improve on this, we next consider a slightly sophisticated, still single-snapshot, model which can approximate these physical effects without the need to solve for the full ionization history. 

To mimic the effect of feedback, we use two characteristic threshold mass \emph{instead of a single one}. Let the minimum mass of ionizing photon producing haloes be $M_{\mathrm{ne}}$ in the neutral regions, while let it be $M_{\mathrm{io}}$ in the ionized regions. The conservation of photons would then lead to the relation
\be
\label{eq:Q_toy_fb}
\left(1-Q_{\mathrm{HII}}^M\right) \left \langle \zeta~f_{\mathrm{coll,i}}(M_{\mathrm{ne}})\Delta_i \right \rangle + Q_{\mathrm{HII}}^M \left \langle \zeta~f_{\mathrm{coll,i}}(M_{\mathrm{io}})\Delta_i \right \rangle = Q_{\mathrm{HII}}^M.
\ee
For given $M_{\mathrm{ne}}$, $M_{\mathrm{io}}$ and $\zeta$, the above can be solved to obtain $Q_{\mathrm{HII}}^M$ without generating the ionization maps. Note that this is applicable only to photon conserving models like ours. 

The above \eqn{eq:Q_toy_fb} can be written equivalently as
\bear
Q_{\mathrm{HII}}^M &= \left(1-Q_{\mathrm{HII}}^M\right) \left[\left \langle \zeta~f_{\mathrm{coll,i}}(M_{\mathrm{ne}})\Delta_i \right \rangle - \left \langle \zeta~f_{\mathrm{coll,i}}(M_{\mathrm{io}})\Delta_i \right \rangle \right]
\nline
& \quad + \left \langle \zeta~f_{\mathrm{coll,i}}(M_{\mathrm{io}})\Delta_i \right \rangle.
\ear
If we further write $Q_{\mathrm{HII}}^M = \left \langle \zeta_{\mathrm{eff}} f_{\mathrm{coll,i}}(M_{\mathrm{ne}})\Delta_i \right \rangle$, this allows us to identify the effective ionization efficiency as
\be
\label{eq:toy_fb}
 \zeta_{\mathrm{eff}}(M) = \begin{cases}
    \zeta, & \text{if $M \ge M_{\mathrm{io}}$},\\
    \left(1-Q_{\mathrm{HII}}^M\right)\zeta, & \text{if $M_{\mathrm{ne}} \le M < M_{\mathrm{io}}$},\\
    0, &\text{otherwise}.
  \end{cases}
\ee
Thus, the implementation of our feedback prescription is equivalent to introducing a halo mass-dependent efficiency having the form above. This is quite simple to implement in the photon conserving algorithm without compromising on the computing efficiency significantly, e.g., see \citet{2022MNRAS.511.2239M} for similar models.

This simple model of the feedback is different from the detailed implementation in several aspects. Firstly, the effect of feedback and hence the effective $M_{\mathrm{io}}$ depends on the thermal history of a cell and thus is not described by a single parameter in the whole box in the detailed model. Further, whether one should use $M_{\mathrm{ne}}$ or $M_{\mathrm{io}}$ in a given cell would depend on whether the cell is already ionized or not. However, in the simple model, we have assumed that a fraction $Q_{\mathrm{HII}}^M$ is determined by $M_{\mathrm{io}}$ and rest by $M_{\mathrm{ne}}$ in \emph{every cell}, irrespective of their ionization state. This needs to be done because we do not know beforehand the ionization state of a cell without generating the ionization map.

The other important effect during reionization is the recombination between ionized atoms and free electrons, which is also intrinsically inhomogeneous in nature. Ideally, one needs to track the density evolution and ionization history of each cell in the simulation box to model this inhomogeneous effect \citep{2013MNRAS.432.3340S,2022MNRAS.511.2239M}. But, this full evolution can be inefficient for a high resolution box which is needed for 21~cm power spectra computation. So, we follow a mean free path-based approach based on \citet{2022MNRAS.514.1302D} where the recombinations can be implemented at a particular redshift without solving for the full ionization evolution.

The implementation requires us to modify \eqn{eq:nion_avail} for computing the available number density of ionizing photons as

\be
\label{eq:mfp_crit}
n_{\mathrm{ion, avail}, i} = \sum_j n_{\mathrm{ion}, j \to i} \longrightarrow \sum_j n_{\mathrm{ion}, j \to i}\left[\frac{x_{ij}}{\lambda (1 - \e^{- x_{ij} / \lambda})}\right]^{-1}
\ee
where $x_{ij}$ is the distance between the cells $i$ and $j$ and $\lambda$ is the mean free path in comoving units. The above modification essentially mimics the loss of photons in a cell from recombination. The effectiveness of the recombinations is characterized by $\lambda$ which is a free parameter in the model. A large $\lambda \to \infty$ would correspond to the model with no recombinations.

There is one more effect we need to account for, which is the photon absorption inside the source cells where they originate. If we naively apply the above equation, these would correspond to $x_{ij} \to 0$ and the effect of recombinations would be absent. Instead, we modify the computation of the number of ionizing photons available in the source cells as \citep{2016MNRAS.460.1328D}

\be
n_{\mathrm{ion}, j} \longrightarrow n_{\mathrm{ion}, j} \left[\frac{\epsilon \Delta x}{\lambda (1 - \e^{- \epsilon \Delta x / \lambda})}\right]^{-1}
\ee
where $\Delta x$ is the cell size and $\epsilon$ is a fudge factor. The value of $\epsilon$ is tuned so that the results are independent of the resolution used. We find the value to be $\epsilon = 0.3$. 

The four-parameter single-snapshot template model thus has four free parameters: $\zeta$, $M_{\mathrm{ne}}$, $M_{\mathrm{io}}$ and $\lambda$. The next step would be to check if the model can be used to recover the reionization history from the 21~cm power spectra.

\subsection{Recovery using the four parameter template model}

\begin{table*}
\renewcommand{\arraystretch}{1.4}
\begin{tabular}{|c|c|c|c|c|c|c|}
\hline
Parameter & Prior & \multicolumn{5}{c}{Mock using full-history, recovery using four parameter single-snapshot model}\\
  
\hline
\multicolumn{3}{|c|}{$z=7.0$}  & \multicolumn{2}{|c|}{$\Delta x=2h^{-1}\mathrm{cMpc}$} & \multicolumn{2}{|c|}{$\Delta x=4h^{-1}\mathrm{cMpc}$} \\
\cline{1-7}
& & input & mean [68$\%$ C.L.] & best-fit  & mean [68$\%$ C.L.] & best-fit \\

\hline
 $\zeta$ &[2, 100] &13.36 & $9.426^{+1.225}_{-1.920}$ & 11.000  & $11.726^{+2.057}_{-5.969}$ & 24.075  \\

 $\log(M_{\mathrm{ne}} / M_{\odot})$ & [7, 9] & 8.15 & $<7.386$ & 7.356  & $<7.509$ & 7.025 \\
$\log(M_{\mathrm{io}} / M_{\odot})$ & [$\log(M_{\mathrm{ne}} / M_{\odot})$, 11] & - & $>10.289$ & 10.978  & $>10.12$ & 10.958 \\

$\log(\lambda / h^{-1}~\mathrm{cMpc})$ & [-2.5,2.5] & - & $1.442^{+0.693}_{-0.654}$ & 0.9940  & $1.229^{+0.745}_{-0.793}$ & 0.245 \\

 $Q_{\mathrm{HII}}^M$ &[0.1,1] & 0.48 & $0.491^{+0.017}_{-0.030}$ & 0.510  & $0.472^{+0.033}_{-0.025}$ & 0.534 \\
 $\chi^2 / \nu$ &$-$ &$-$ & $-$ & $4.383/14$ & $-$   &  $2.485/6$ \\

\hline
\multicolumn{3}{|c|}{$z=8.0$}  & \multicolumn{2}{|c|}{$\Delta x=2h^{-1}\mathrm{cMpc}$} & \multicolumn{2}{|c|}{$\Delta x=4h^{-1}\mathrm{cMpc}$} \\
\hline
 $\zeta$ &[2, 100] &10.19 & $< 18.809$ & 10.420  &$< 32.319$ &21.212 \\

$\log(M_{\mathrm{ne}} / M_{\odot})$ & [7, 9] & 8.07 & $< 7.316$ & 7.381  & $7.892^{+0.469}_{-0.215}$ & 7.378 \\
$\log(M_{\mathrm{io}} / M_{\odot})$ & [$\log(M_{\mathrm{ne}} / M_{\odot})$, 11] & - & $>10.223$ & 10.970  & $> 9.874$ & 9.247 \\

$\log(\lambda / h^{-1}~\mathrm{cMpc})$ & [-2.5, 2.5] & - & $ 0.394^{+0.678}_{-1.261}$ & -0.196  & $0.392^{+0.392}_{-1.157}$ & -0.307 \\
 $Q_{\mathrm{HII}}^M$ & [0.1,1] & 0.27 & $0.283^{+0.005}_{-0.005}$ & 0.278  & $0.265^{+0.010}_{-0.020}$ & 0.259\\
 $\chi^2 / \nu$ &$-$ &$-$ & $-$ & $4.397/14$ & $-$   &$2.642/6$ \\
\hline
\multicolumn{3}{|c|}{$z=6.2$}  & \multicolumn{2}{|c|}{$\Delta x=2h^{-1}\mathrm{cMpc}$} & \multicolumn{2}{|c|}{$\Delta x=4h^{-1}\mathrm{cMpc}$} \\
\hline
 $\zeta$ &[2, 100] & 17.03 & $30.767^{+5.916}_{-14.854}$ & 37.488 &$32.503^{+7.206}_{-16.664}$ & 37.328\\

 $\log(M_{\mathrm{ne}} / M_{\odot})$ & [7, 9] & 8.21 & $7.381^{+0.226}_{-0.149}$ & 7.638  & $7.330^{+0.209}_{-0.190}$ & 7.574 \\
$\log(M_{\mathrm{io}} / M_{\odot})$ & [$\log(M_{\mathrm{ne}} / M_{\odot})$, 11] & - & $>10.724$ & 10.995  & $>10.725$ & 10.997 \\

$\log(\lambda / h^{-1}~\mathrm{cMpc})$ & [-2.5, 2.5] & - & $1.472^{+0.335}_{-0.563}$ & 1.202  & $1.425^{+0.282}_{-0.545}$ & 1.182 \\
 $Q_{\mathrm{HII}}^M$ &[0.1,1] & 0.77 & $0.751^{+0.016}_{-0.011}$ & 0.765  & $0.745^{+0.020}_{-0.013}$ & 0.764\\
 $\chi^2 / \nu$ &$-$ &$-$ & $-$ & 6.245/14 & $-$   &  $6.009/6$ \\
\hline 
\end{tabular}
\caption{Parameter constraints obtained using the photon conserving four parameter single-snapshot model at redshifts $z=7.0,~8.0$ and $6.2$. Results are shown for two different resolutions $\Delta x = 2 h^{-1}\mathrm{cMpc}$ and $4 h^{-1}\mathrm{cMpc}$ used for the MCMC analysis. The mock data is generated using the full-history model which includes radiative feedback and inhomogeneous recombinations, always at the resolution $\Delta x = 2 h^{-1}\mathrm{cMpc}$. For each parameter, we show the prior and the input value used for the mock (wherever possible) along with the obtained mean, $68\%$ confidence limits and best-fit. We also provide the $\chi^2 / \nu$ for the best-fit models, $\nu$ being the number of degrees of freedom. 
}
\label{tab:table_2}
\end{table*}

\begin{figure}
    \includegraphics[width=0.5\textwidth]{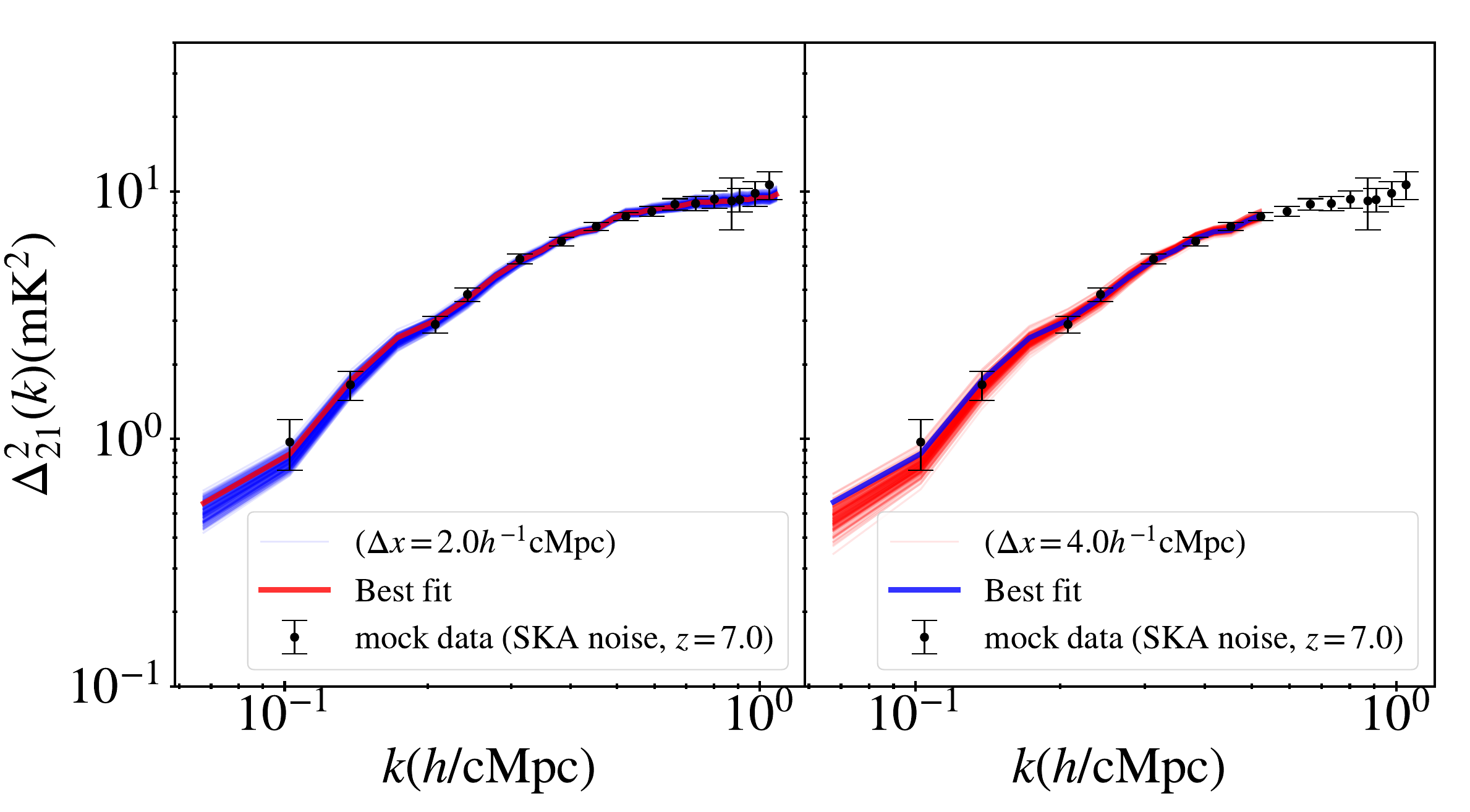}
    \caption{The 21~cm power spectra for the photon conserving four parameter single-snapshot model at redshift $z=7$ for 200 random samples drawn from the posterior distributions shown in \fig{fig:toy_recom_feed_z7}. The \textit{blue} lines show the models for a resolution $\Delta x=2 h^{-1} \mathrm{cMpc}$ while the \textit{red} ones are for $\Delta x=4h^{-1}\mathrm{cMpc}$. The dashed lines denote the input models.}
    \label{fig:toy_recom_feed_pow_7}
\end{figure}

\begin{figure*}
    \includegraphics[width=0.95\textwidth]{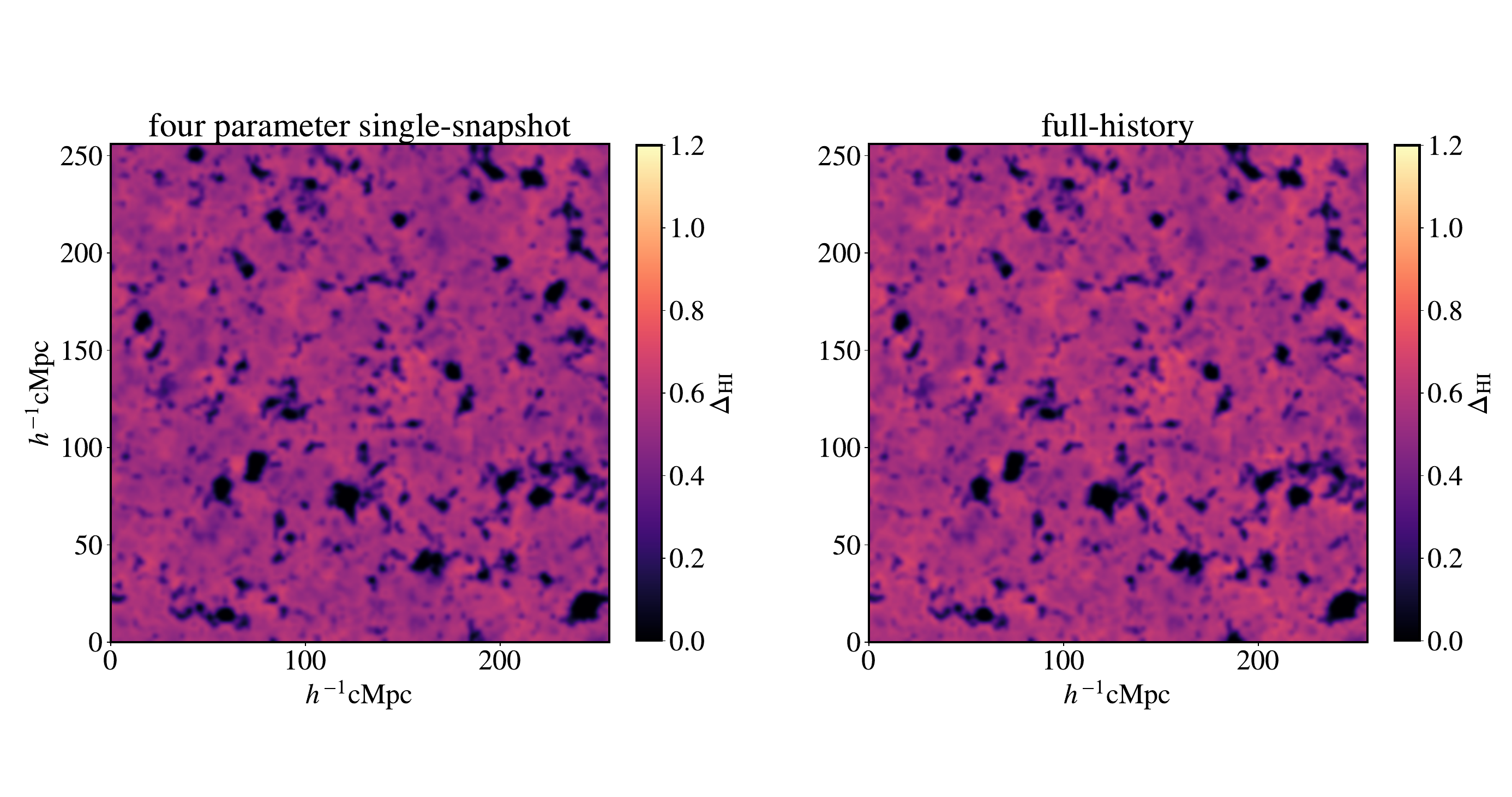}
    \caption{A comparison of ionization maps between the four parameter single-snapshot model (left) and the full-history model (right) for $z = 7$. The quantity plotted is $\Delta_{\mathrm{HI}, i} = (1 - x_{\mathrm{HII}},i)~\Delta_i$ which essentially determines the flucuations in the 21~cm signal, see \eqn{eq:delta_Tb}. The parameter values used for the single-snapshot model corresponds to the best-fits obtained in the MCMC chains.}
    \label{fig:toy_model_ionization_map_7}
\end{figure*}
\begin{figure}
    \centering
    \includegraphics[width=0.99\columnwidth]{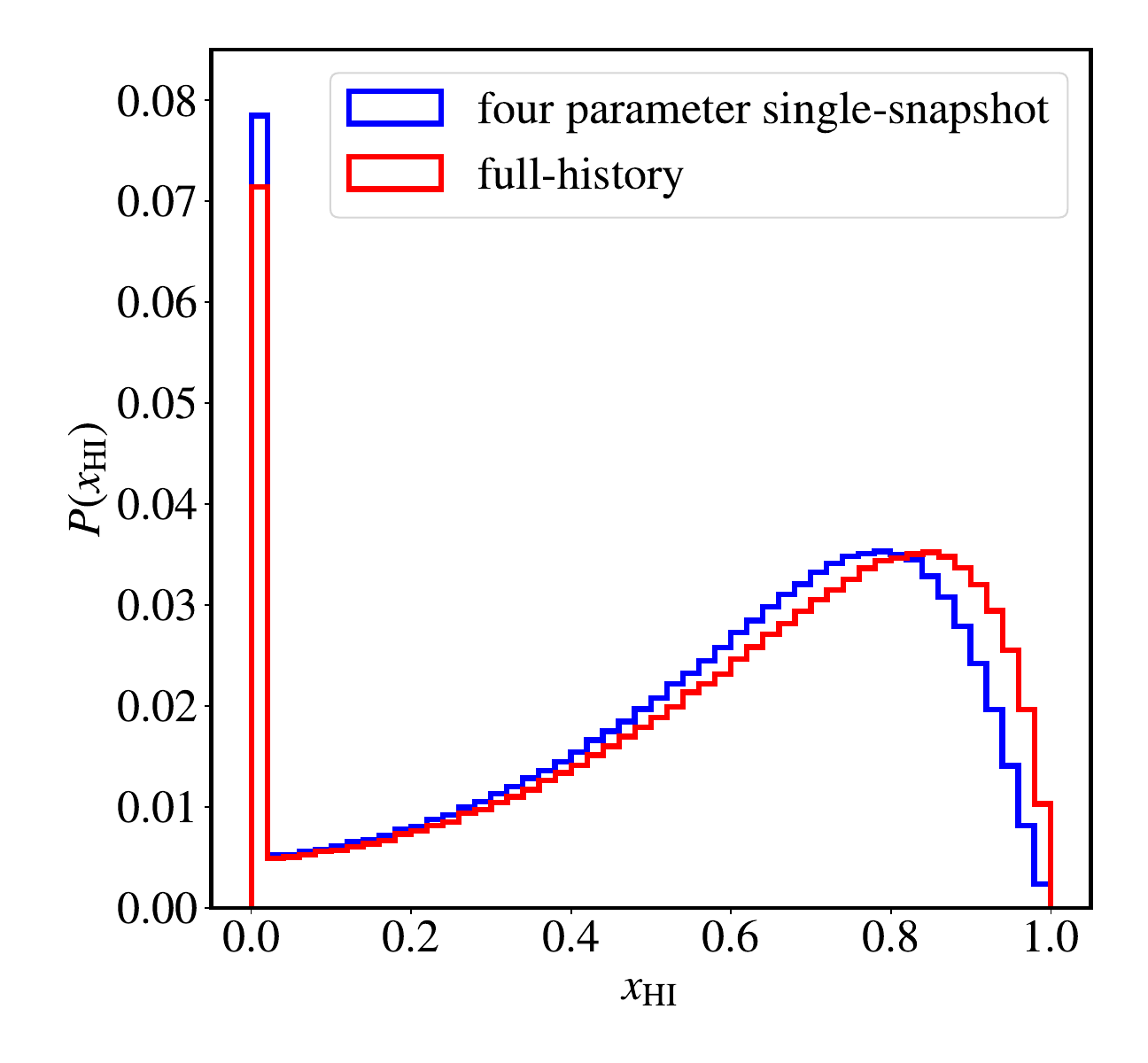}
    \caption{The probability distribution of neutral fraction $x_{\mathrm{HI}} \equiv 1 - x_{\mathrm{HII}}$ in the box for the four parameter single-snapshot model and full-history model at $z = 7$. The parameter values used for the single-snapshot model corresponds to the best-fits obtained in the MCMC chains.}
    \label{fig:neu_pdf}
\end{figure}

We now present the results when the mock data is generated using the full-history model of \secn{sec:full_history} while the Bayesian analysis is carried out using the single-snapshot template introduced above. For the parameters $\zeta$, $M_{\mathrm{ne}}$ and $\lambda$, we choose flat priors over a sufficiently wide range. We also impose a physically motivated condition on $M_{\mathrm{io}}$ that it should be larger than $10^9 M_{\odot}$ and also that $M_{\mathrm{io}} \ge M_{\mathrm{ne}}$ \citep{2013MNRAS.432.3340S,2021MNRAS.503.3698H}. 

Let us begin our discussion by showing the results at $z = 7$ (the midpoint of reionization). \fig{fig:toy_recom_feed_z7} shows the posteriors of the model parameters, the corresponding values are quoted in \tab{tab:table_2}. Usually, the quality of parameter recovery is assessed by comparing the recovered values with the inputs. However, in this case, there is no input counterpart of the two parameters $M_{\mathrm{io}}$ and $\lambda$. The input values of the other parameters, including the derived $Q_{\mathrm{HII}}^M$, are marked by dashed lines in the figure. We can see the recovered values of both $\zeta$ and $M_{\mathrm{ne}}$ are lower than the input value for the resolution $\Delta x = 2 h^{-1}$~cMpc corresponding to the one used for generating the mock data (blue curves and regions). This clearly suggests that the simplistic four-parameter single-snapshot model cannot recover the source properties faithfully. The lower recovered $\zeta$ implies that the recombinations could be underestimated in the simple mean free path approach of the single-snapshot model. A low $\zeta$ in the model is compensated by a $M_{\mathrm{ne}}$ higher than the input value used in the mock data. What is interesting is that, in spite of the biased estimates of the source properties, the recovered value of $Q_{\mathrm{HII}}^M$ is a remarkable match to the input value. This result points towards one utility of the single-snapshot model: it can be used to recover the reionization history faithfully. \fig{fig:toy_recom_feed_pow_7} shows that the match between the mock power spectra and the recovered ones are also excellent.

The argument can be strengthened by carrying out the analysis with a different resolution $\Delta x = 4 h^{-1}$~cMpc (keeping the mock data same, i.e., generated at a fine resolution of $\Delta x = 2 h^{-1}$~cMpc). The results are shown in \figs{fig:toy_recom_feed_z7} and \ref{fig:toy_recom_feed_pow_7} by the red curves and contours. Firstly, the posteriors of all the parameters are consistent between the two resolutions, thus confirming the numerical convergence of the four parameter single-snapshot model. More importantly, there is no bias in the recovery of $Q_{\mathrm{HII}}^M$ for this resolution as well.

We can test our postulate that the single-snapshot template model provides a good description of the HI field by comparing the ionization maps visually. The maps are shown in \fig{fig:toy_model_ionization_map_7} where we plot the quantity $\Delta_{\mathrm{HI},i} = (1 - x_{\mathrm{HII}, i})~\Delta_i$. One can see clearly that the maps are very similar. In \fig{fig:neu_pdf}, we also show the probability distribution function of neutral fraction $1 - x_{\mathrm{HII}}$ for the full-history model and the four parameter single-snapshot model. It is evident that most of the cells are ionized and rest of the cells are at a low ionized state for both the cases. This is in agreement with the expected trend, i.e., most of the grid cells are at a fully ionized state or at a highly neutral state. However, there also exist a significant fraction of cells that are partially ionized, this is a consequence of our grid cells being relatively coarse. What is important for our work is that the two models produce distributions which agree quite well. This indicates that the simplistic models can be useful for characterizing the properties of the IGM, even though they are limited in recovering the source properties. Interestingly, using the present limits on the 21~cm power spectrum, it has become possible to put constraints on the properties of the ionized (and hence heated) regions \citep{2020MNRAS.493.4728G}; the simple model can be useful in this regard.

Let us next study the recovery of the parameters at other redshifts. We show the parameter constraints in \tab{tab:table_2} for $z = 8$ and $6.2$. The recovery of $Q_{\mathrm{HII}}^M$ can be seen to be reasonable. The worst case seems to be for $z = 8$ when the analysis is carried out at the finer resolution. Because of tiny error-bars on the parameter, the input value is outside the $2 \sigma$ region of the recovered constraints. However, even in this case, the difference in the recovered and input values are within 5 per cent. We can thus hope to recover the reionization history reliably using the single-snapshot template model, the advantage being that the analysis requires very little computational resources. At the same time, we stress that the single-snapshot models are unlikely to be reliable for understanding the properties of the reionization sources, e.g., the ionizing photon efficiency and the mass of the haloes hosting the ionizing sources. This limitation should be kept in mind while using these models to interpret observations.

Although the analysis of this section holds a promise that one can obtain the reionization history from 21~cm maps using a rather simple model of radiative feedback and inhomogeneous recombinations, one should keep in mind that the underlying procedure for generating the ionization maps is almost identical for the mock data and the model used to constrain the parameters. It is thus not clear whether using widely different algorithms for ionization maps would lead to similar conclusions regarding the value of $Q_{\mathrm{HII}}^M$. In fact, we discuss in \appndx{app:ES_PC} that two different methods of generating ionization maps, namely excursion set and photon conserving, could lead to very different conclusions on $Q_{\mathrm{HII}}^M$ even when the source model is exactly the same. At this point, we can only claim that the simple single-snapshot model provides an initial direction on how to obtain the reionization history quickly from the high quality data expected in the future. Another way to appreciate the importance of the analysis is that it is a test case where the theoretical model is only an approximate description of the universe which provided the observational data. Our speculation is that we can recover the reionization history and the IGM properties with the approximate model, but the source properties require way more detailed modelling and/or comparing with other observations.

\section{Summary \& Conclusions}
\label{sec:conc}

The 21~cm power spectrum of neutral hydrogen from the epoch of reionization is a promising means for constraining the effect of the first stars on the ionization history of the universe. The parameters characterizing the physical processes during this epoch can be constrained by comparing the observational data with theoretical models. In case one attempts to obtain the constraints using MCMC based Bayesian inference techniques, it requires the model to be evaluated many times for different parameter values. As a result, these models need to be computationally efficient. In this work, we check the prospects of constraining the reionization parameters using the photon conserving semi-numerical model \texttt{SCRIPT}. Our main aim is to ensure that the model provides unbiased estimates of the underlying parameters, irrespective of the resolution of the simulation used. For this purpose, we generate and use mock data sets as expected from the upcoming SKA-Low in $\sim 1000$ hours of observations.

The main results of our work can be summarized as follows:

\begin{itemize}

\item For the simplest model where the reionization can be modelled using only two parameters, namely, the ionizing photon efficiency $\zeta$ and the minimum threshold halo mass $M_{\mathrm{min}}$ that can contribute to ionizing photons, we find that our model can recover the input parameters extremely well. The conclusion holds for different phases of reionization (and hence different redshifts)  and also for different resolutions of the simulation. In particular, the model provides unbiased estimates of the parameters even when the resolution of the simulation used for parameter constraints is different from that used to generate the mock data. We find that the same conclusion does not hold for excursion set based models for generating ionization maps.

\item We also explore the capability of our model to recover parameters when the reionization model is taken to be more complex, e.g., by including radiative feedback and inhomogeneous recombinations. This requires us to compute the ionization and thermal histories consistently, and hence the model becomes computationally slower. So, for parameter recovery, we develop a simple template model where these physical processes are approximated by additional parameters. The advantage of this approximate model is that the ionization maps can be computed at a single redshift without any reference to the overall history. In this case, we find that the template single-snapshot model cannot recover the source properties faithfully, which clearly indicates the limitations of such models. However, it manages to recover the global ionization fraction without any bias, over the full reionization history and for different resolutions. This promises that the template can be used to recover the reionization history with moderate computational resources.

\end{itemize}

The work highlights the necessity of photon number conservation while interpreting the 21~cm data. The large scale convergence with respect to the resolution is important as we do not know about the correct resolution to work with when real observational data will be available. So, it becomes important to take into account the photon number conservation while creating the ionization maps.

There are studies \citep{2019MNRAS.484..933P,2021MNRAS.506.2390Q} which have shown that the constraints on the reionization parameters improve significantly after including 21~cm data with measurements from other observational probes. So, the next obvious target is to study the joint estimates using the realistic 21~cm mocks and the observational data used in \citet{2022MNRAS.515..617M}. These require a more efficient way to compute the models with full ionization history using a high resolution. In a future project, we are planning to build up an emulator which can serve the purpose. These kinds of studies will be very useful before the real observational data become available.

\section*{Acknowledgements}

The authors acknowledge support of the Department of Atomic Energy, Government of India, under project no. 12-R\&D-TFR-5.02-0700.

\section*{Data availability}

A basic version of the code, which does not include the effects of recombinations and feedback on ionization maps, used in the paper is publicly available at \url{https://bitbucket.org/rctirthankar/script}. The data obtained from the extensions of the code and presented in this article will be shared on reasonable request to the corresponding author (BM).

\bibliographystyle{mnras}
\bibliography{21cm_script} 

\appendix
\section{Recovery using Excursion set Based Approach}
\label{app:es_based}

\begin{figure}
    \includegraphics[width=0.99\columnwidth]{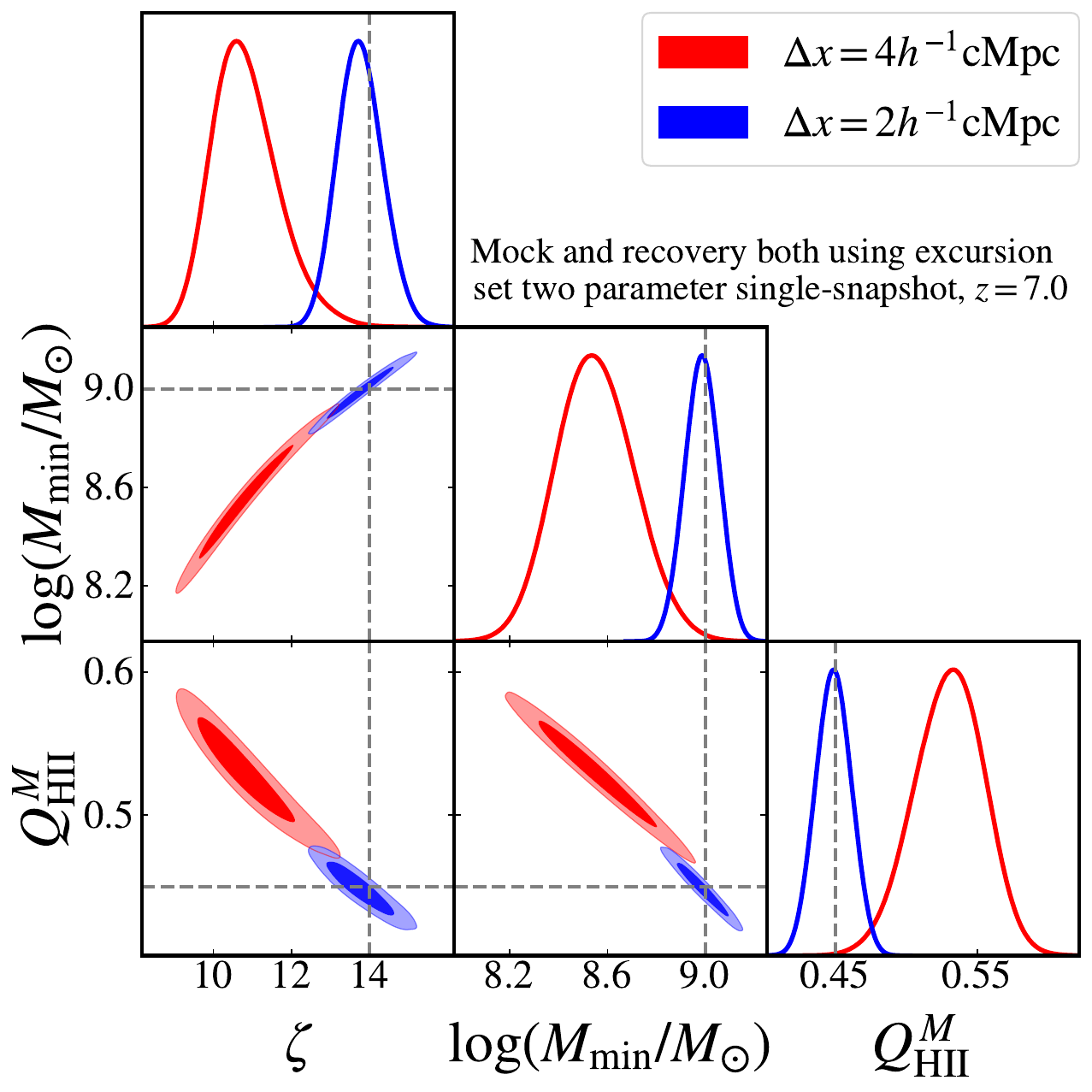}
    \caption{Recovery of parameters using the excursion set two parameter single-snapshot model at redshift $z=7$. The mock data is created by the same single-snapshot model (but with a different realization of the matter density field) using a grid resolution $\Delta x = 2 h^{-1}$~cMpc. The off diagonal panels show the joint two-dimensional posterior distributions of each pair of parameters. The contours represent $68\%$ and $95\%$ confidence intervals. The diagonal panels show the marginalized posterior distributions of the parameters. The dashed lines represent the input values for generating mock data. The \textit{blue} contours and curves show the results when the resolution used for the MCMC run is $\Delta x = 2 h^{-1}$~cMpc (same as the one used for the mock data), while the \textit{red} ones are for the resolution $\Delta x = 4 h^{-1}$~cMpc. The recovery of the input parameters are biased when the resolution used is different from that used to generate the mock.}
    \label{fig:ES_two_params_7}
\end{figure}

\begin{figure}
    \includegraphics[width=0.5\textwidth]{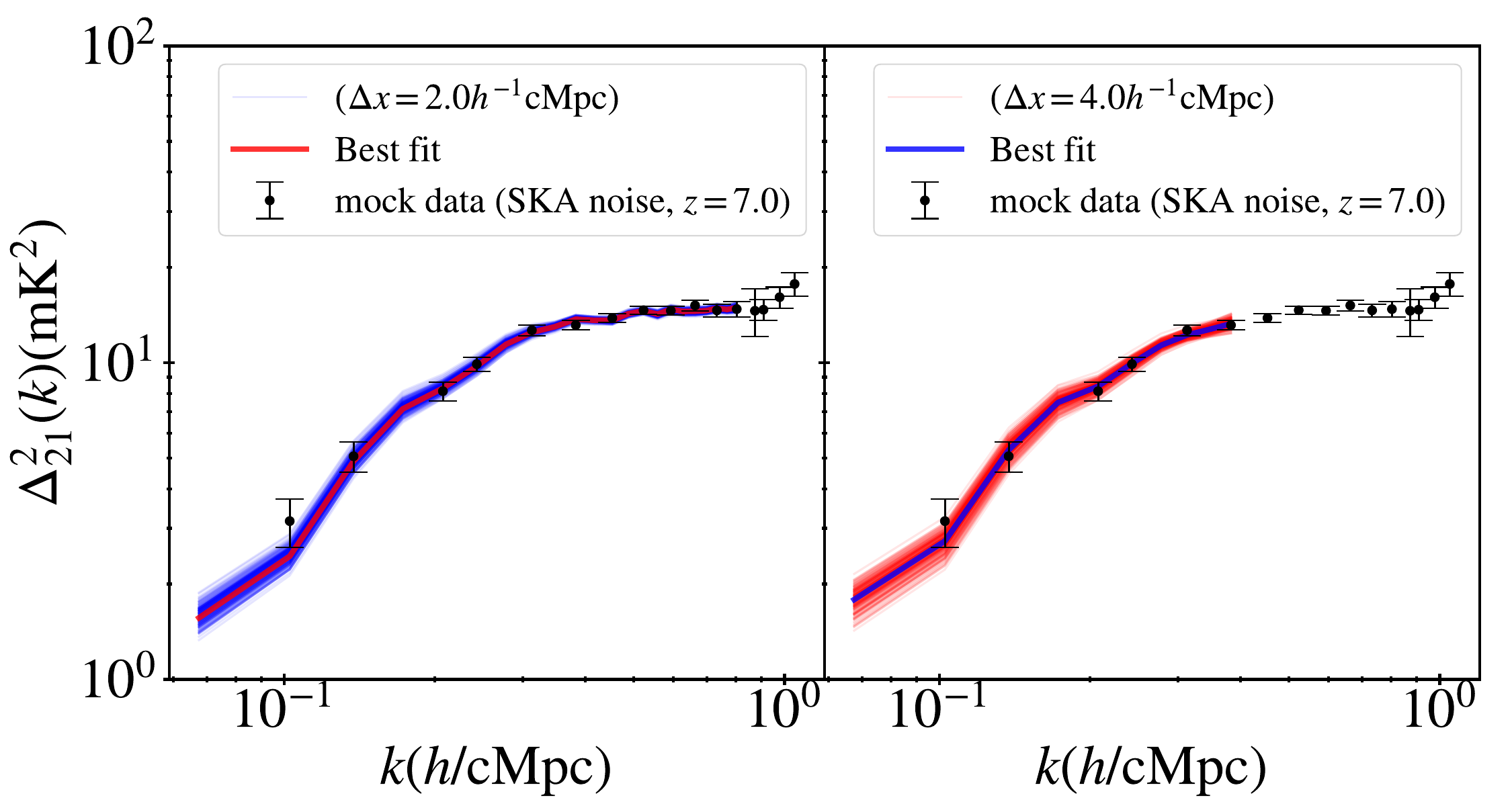}
    \caption{The 21~cm power spectra for the excursion set two parameter single-snapshot model at redshift $z=7$ for 200 random samples drawn from the posterior distributions shown in \fig{fig:ES_two_params_7}. The \textit{blue} lines show the models for a resolution $\Delta x=2 h^{-1} \mathrm{cMpc}$ while the \textit{red} ones are for $\Delta x=4h^{-1}\mathrm{cMpc}$. The dashed lines denote the input models.}
    \label{fig:ES_two_params_pow_7}
\end{figure}

In \secn{sec:two_param_model}, we discussed the recovery of model parameters for a simple two parameter model using our photon conserving code \texttt{SCRIPT}. In this appendix, we discuss an identical analysis on parameter recoveries but for the excursion set based model. It is known that the excursion set based models do not conserve the number of photons  \citep{2007ApJ...654...12Z, 2011MNRAS.414..727Z,2016MNRAS.460.1801P}. It was shown in an earlier work by \citet{2018MNRAS.481.3821C} that an important consequence of this non-conservation is the non-convergence of the large scale 21~cm power spectra for different resolutions. The aim of the analysis here is to check how the parameter recoveries are affected by the non-convergent power spectra.

Similar to the previous case, we generate the mock dataset for the finest resolution ($\Delta x =2 h^{-1} \mathrm{cMpc}$) using the excursion set based model. The density and collapsed halo fields used for this purpose are identical to those used for the photon conserving case. The details of our implementation of the excursion set model can be found in earlier papers \citep{2009MNRAS.394..960C,2014MNRAS.443.2843M}, in particular see Section 3.1 of \citet{2018MNRAS.481.3821C}. We calculate the uncertainties due to the thermal noise of the telescope and the cosmic variance to the mock power spectrum as in \eqn{eq:tot_noise}. We then shift the power spectrum data points by adding a random number having a Gaussian distribution with zero mean and a standard deviation equal to the associated thermal noise.

Let us discuss the case which corresponds to the middle stages of the reionization, i.e., $z=7$.  We use parameters $\zeta = 14$ and $M_{\mathrm{min}} = 10^9 \Msun$ as the input model to generate the mock data. The global ionization fraction for this input model is $Q_{\mathrm{HII}}^M = 0.45$. We first study the recovery when the resolution is the same $\Delta x = 2 h^{-1} \mathrm{cMpc}$ as the one used for generating the mock data. The posterior distributions are shown in blue in \fig{fig:ES_two_params_7}, the corresponding values can be found in \tab{tab:table_es}. It is clear that the recovery of all the parameters is excellent when the resolution used for the analysis is the same as the mock data.

\begin{table*}
\renewcommand{\arraystretch}{1.4}
\begin{tabular}{|c|c|c|c|c|c|c|}
\hline
Parameter & Prior & \multicolumn{5}{c}{Excursion set two parameter single-snapshot model}\\ %
  
\hline
\multicolumn{3}{|c|}{$z=7.0$}  & \multicolumn{2}{|c|}{$\Delta x=2h^{-1}\mathrm{cMpc}$} & \multicolumn{2}{|c|}{$\Delta x=4h^{-1}\mathrm{cMpc}$} \\
\cline{1-7}
& & input & mean [68$\%$ C.L.] & best-fit  & mean [68$\%$ C.L.] & best-fit \\

\hline
  $\zeta$ &[2, 100] &14 & $13.774^{+0.545}_{-0.605}$ & 13.827  & $10.830^{+0.678}_{-0.956}$ & 10.668  \\

 $\log(M_{\mathrm{min}} / M_{\odot})$ & [7, 11] & 9 & $8.988^{+0.070}_{-0.071}$ & 9.000  & $8.550^{+0.153}_{-0.167}$ & 8.531\\
 $Q_{\mathrm{HII}}^M$ &[0.1,1] & 0.45 & $0.448^{+0.012}_{-0.013}$ & 0.4446  & $0.529^{+0.026}_{-0.023}$ & 0.533\\
 $\chi^2 / \nu$ &$-$ &$-$ & $-$ & $5.613/10$ & $-$   &  $1.899/4$ \\

\hline
\multicolumn{3}{|c|}{$z=8.0$}  & \multicolumn{2}{|c|}{$\Delta x=2h^{-1}\mathrm{cMpc}$} & \multicolumn{2}{|c|}{$\Delta x=4h^{-1}\mathrm{cMpc}$} \\
\hline
 $\zeta$ &[2, 100] &14 & $11.94^{+2.16}_{-2.19}$ & 12.84 &$8.96^{+2.19}_{-2.32}$ & 10.23 \\

 $\log(M_{\mathrm{min}} / M_{\odot})$ & [7, 11] & 9 & $8.79^{+0.33}_{-0.18}$ & 8.90  & $8.49^{+0.62}_{-0.27}$ & 8.89 \\
 $Q_{\mathrm{HII}}^M$ & [0.1,1] & 0.26 & $0.28^{+0.03}_{-0.05}$ & 0.27017  & $0.28^{+0.04}_{-0.08}$ & 0.23\\
 $\chi^2 / \nu$ &$-$ &$-$ & $-$ & $11.8998/10$ & $-$   &$6.6739/4$ \\
\hline
\multicolumn{3}{|c|}{$z=6.2$}  & \multicolumn{2}{|c|}{$\Delta x=2h^{-1}\mathrm{cMpc}$} & \multicolumn{2}{|c|}{$\Delta x=4h^{-1}\mathrm{cMpc}$} \\
\hline
 $\zeta$ &[2, 100] &14 & $14.98^{+1.08}_{-2.06}$ & 14.18 &$15.24^{+0.69}_{-1.11}$ & 15.36\\

 $\log(M_{\mathrm{min}} / M_{\odot})$ & [7, 11] & 9 & $9.05^{+0.08}_{-0.13}$ & 9.01 & $9.14^{+0.05}_{-0.06}$ & 9.13\\
 $Q_{\mathrm{HII}}^M$ &[0.1,1] & 0.71 & $0.72^{+0.02}_{-0.02}$ & 0.71  & $0.68^{+0.02}_{-0.02}$ & 0.69\\
 $\chi^2 / \nu$ &$-$ &$-$ & $-$ & $11.7213/10$ & $-$   &  $15.2057/4$ \\
\hline 
\end{tabular}
\caption{Parameter constraints obtained using the excursion set two parameter single-snapshot model at redshifts $z=7.0,~8.0$ and $6.2$. Results are shown for two different resolutions $\Delta x = 2 h^{-1}\mathrm{cMpc}$ and $4 h^{-1}\mathrm{cMpc}$ used for the MCMC analysis. The mock data is generated using the same excursion set model, always at the resolution $\Delta x = 2 h^{-1}\mathrm{cMpc}$. For each parameter, we show the prior and the input value used for the mock along with the obtained mean, $68\%$ confidence limits and best-fit. We also provide the $\chi^2 / \nu$ for the best-fit models, $\nu$ being the number of degrees of freedom. 
}
\label{tab:table_es}
\end{table*}

However, the posterior distributions for the coarser resolution $\Delta x = 4 h^{-1}$~cMpc (red) show significant deviation from the finer ones. The input values (shown by dashed lines) deviate significantly from the two-dimensional joint posterior distributions. The same can be concluded from \tab{tab:table_es} too. If we concentrate on $Q_{\mathrm{HII}}^M$, we find that the best-fit values for the two resolutions differ by $\sim 20\%$ (the corresponding difference was $\sim 2.5\%$ for the photon conserving model, see \tab{tab:table_1}). The recovered mean values of $Q_{\mathrm{HII}}^M$ are inconsistent at a level $\gtrsim 3 \sigma$ (they were well within $1 \sigma$ for the photon conserving case). Thus, using a coarser resolution map for parameter estimation could lead to a significantly biased value of $Q_{\mathrm{HII}}^M$ for the excursion set models.

We see interesting trends for other redshifts as well, see \tab{tab:table_es} for the constraints and compare with the input values. At $z = 8$, we find that the mean values of $Q_{\mathrm{HII}}^M$ for the two resolutions are consistent with each other. This result is consistent with the findings in our earlier work \citep{2018MNRAS.481.3821C} that the effect of photon non-conservation is less at early stages of reionization when the bubble sizes are smaller than the grid resolution used. At late stages of reionization $z = 6.2$, the mean values of $Q_{\mathrm{HII}}^M$ for the two resolutions are within $\sim 1.5 \sigma$, which shows that the effect is less at late stages too. The effect of photon non-conservation, leading to non-convergence of the power spectra, is maximum at the middle stages of reionization where the large-scale signal is maximum.

We should mention here that the results presented here are valid \emph{only} for the excursion set model as implemented by us. For other implementations of the algorithm, which could vary from ours regarding how the haloes are identified, smoothing of the density field, type of filters used to identify self-ionized regions, the results could be different. However, the main lesson from our analysis is that regardless of the implementation, it is important to check for the numerical convergence of the results for any excursion set algorithm.

\section{Recovery using ES model with PC mocks}
\label{app:ES_PC}

\begin{figure}
    \centering
    \includegraphics[width=0.99\columnwidth]{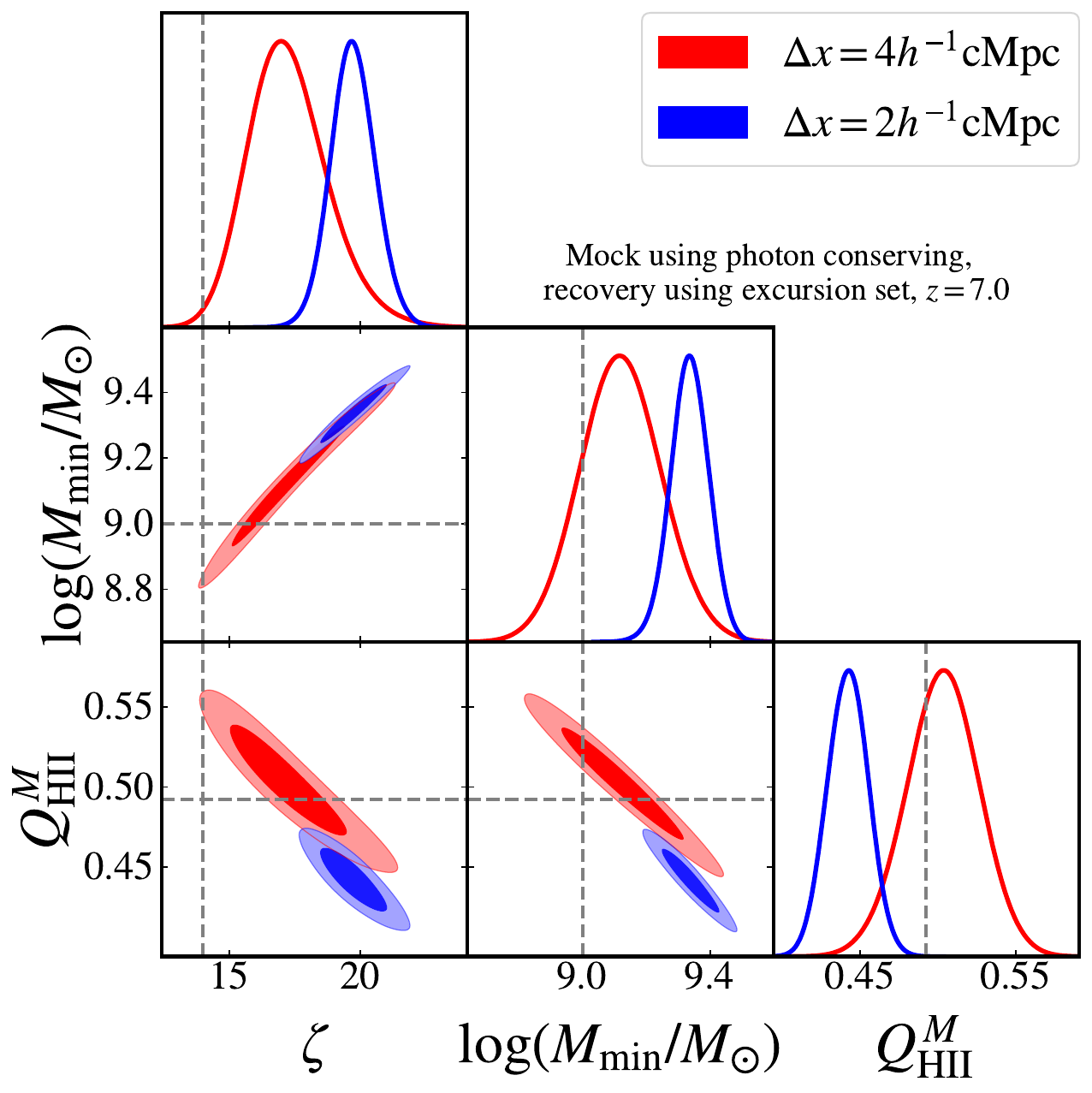}
    \caption{Recovery of parameters using the excursion set two parameter single-snapshot model at redshift $z=7$. The mock data is created by the photon conserving single-snapshot model (with a different realization of the matter density field) using a grid resolution $\Delta x = 2 h^{-1}$~cMpc. The off diagonal panels show the joint two-dimensional posterior distributions of each pair of parameters. The contours represent $68\%$ and $95\%$ confidence intervals. The diagonal panels show the marginalized posterior distributions of the parameters. The dashed lines represent the input values for generating mock data. The \textit{blue} contours and curves show the results when the resolution used for the MCMC run is $\Delta x = 2 h^{-1}$~cMpc (same as the one used for the mock data), while the \textit{red} ones are for the resolution $\Delta x = 4 h^{-1}$~cMpc.}
    \label{fig:PC_mock_ES}
\end{figure}

\begin{figure}
    \centering
    \includegraphics[width=0.5\textwidth]{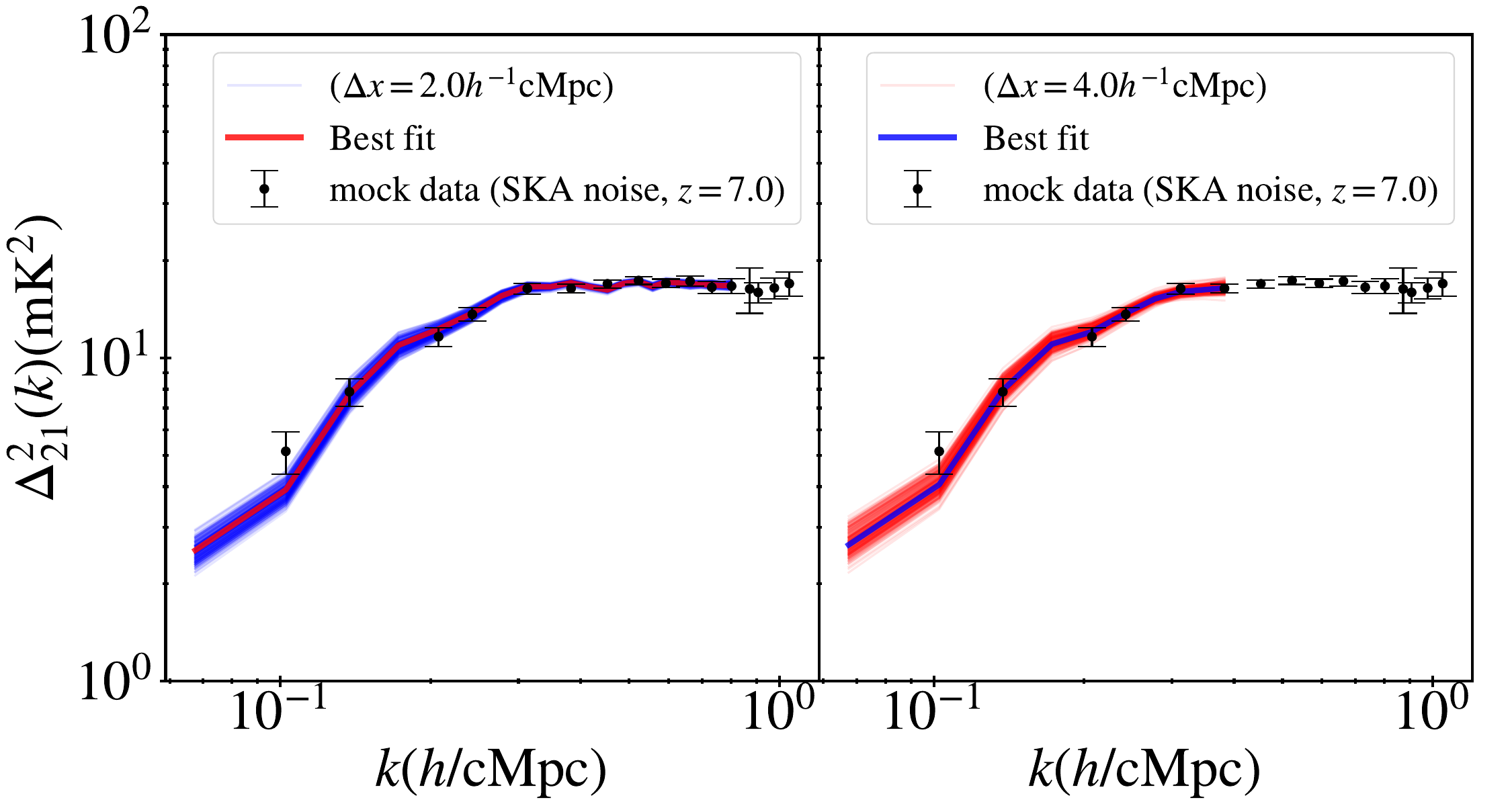}
    \caption{The 21~cm power spectra for the excursion set two parameter single-snapshot model at redshift $z=7$ for 200 random samples drawn from the posterior distributions shown in \fig{fig:ES_two_params_7}. The \textit{blue} lines show the models for a resolution $\Delta x=2 h^{-1} \mathrm{cMpc}$ while the \textit{red} ones are for $\Delta x=4h^{-1}\mathrm{cMpc}$. The dashed lines denote the input models.}
    \label{fig:PC_mock_ES_model}
\end{figure}

\begin{table*}
\renewcommand{\arraystretch}{1.4}
\begin{tabular}{|c|c|c|c|c|c|c|}
\hline
Parameter & Prior & \multicolumn{5}{c}{Mock using photon conserving, recovery using excursion set model}\\ %
  
\hline
\multicolumn{3}{|c|}{$z=7.0$}  & \multicolumn{2}{|c|}{$\Delta x=2h^{-1}\mathrm{cMpc}$} & \multicolumn{2}{|c|}{$\Delta x=4h^{-1}\mathrm{cMpc}$}  \\
\cline{1-7}
& & input & mean [68$\%$ C.L.] & best-fit  & mean [68$\%$ C.L.] & best-fit  \\

\hline
 $\zeta$ &[2, 100] &14 & $19.595^{+0.920}_{-0.853}$ & 19.571 & $17.230^{+1.215}_{-1.649}$ & 17.023  \\

 $\log(M_{\mathrm{min}} / M_{\odot})$ & [7, 11] & 9 & $9.327^{+0.067}_{-0.059}$ & 9.326 & $9.120^{+0.125}_{-0.124}$ & 9.113 \\
 $Q_{\mathrm{HII}}^M$ &[0.1,1] & 0.49 & $0.444^{+0.013}_{-0.014}$ & 0.445 & $0.503^{+0.022}_{-0.022}$& 0.503 \\
 $\chi^2 / \nu$ &$-$ &$-$ & $-$ & $8.053/10$ & $-$& $2.641/4$\\

\hline

\multicolumn{3}{|c|}{$z=8.0$}  & \multicolumn{2}{|c|}{$\Delta x=2h^{-1}\mathrm{cMpc}$}& \multicolumn{2}{|c|}{$\Delta x=4h^{-1}\mathrm{cMpc}$}  \\
\hline
 $\zeta$ &[2, 100] &14 & $18.475^{+2.022}_{-1.949}$ & 18.609 & $20.981^{+9.516}_{-8.925}$& 14.715\\

 $\log(M_{\mathrm{min}} / M_{\odot})$ & [7, 11] & 9 & $9.096^{+0.168}_{-0.127}$ & 9.118 & $9.290^{+0.561}_{-0.466}$ & 8.911\\
 $Q_{\mathrm{HII}}^M$ &[0.1,1] & 0.28 & $0.300^{+0.021}_{-0.025}$ & 0.297 & $0.261^{+0.072}_{-0.082}$ & 0.319\\
 $\chi^2 / \nu$ &$-$ &$-$ & $-$ & $4.346/10$ & $-$& $3.574/4$\\
\hline 

\multicolumn{3}{|c|}{$z=6.2$}  & \multicolumn{2}{|c|}{$\Delta x=2h^{-1}\mathrm{cMpc}$}& \multicolumn{2}{|c|}{$\Delta x=4h^{-1}\mathrm{cMpc}$}  \\
\hline
 $\zeta$ &[2, 100] &14 & $13.756^{+0.714}_{-0.969}$ & 13.702 & $14.500^{+0.550}_{-0.785}$& 14.173\\

 $\log(M_{\mathrm{min}} / M_{\odot})$ & [7, 11] & 9 & $8.968^{+0.062}_{-0.073}$ & 8.965 & $9.089^{+0.050}_{-0.050}$ & 9.071\\
 $Q_{\mathrm{HII}}^M$ &[0.1,1] & 0.75 & $0.716^{+0.009}_{-0.009}$ & 0.717 & $0.680^{+0.012}_{-0.016}$ & 0.676\\
 $\chi^2 / \nu$ &$-$ &$-$ & $-$ & $19.11/10$ & $-$& $9.879/4$\\
\hline 

\end{tabular}
\caption{Parameter constraints obtained using the excursion set two parameter single-snapshot model at redshifts $z=7.0,~8.0$ and $6.2$. Results are shown for two different resolutions $\Delta x = 2 h^{-1}\mathrm{cMpc}$ and $4 h^{-1}\mathrm{cMpc}$ used for the MCMC analysis. The mock data is generated using the two parameter photon conserving model, always at the resolution $\Delta x = 2 h^{-1}\mathrm{cMpc}$. For each parameter, we show the prior and the input value used for the mock along with the obtained mean, $68\%$ confidence limits and best-fit. We also provide the $\chi^2 / \nu$ for the best-fit models, $\nu$ being the number of degrees of freedom. 
}
\label{tab:table_mock_pc_es}
\end{table*}

It has been shown by \citet{2018MNRAS.481.3821C} that the ionization maps and power spectra predicted by the photon conserving and excursion set models, for the same input parameters, differ from each other. To study the consequence of this difference on recovery of parameters, we carry out an exercise where we generate the mock data using the photon conserving model and run the MCMC chains to constrain the parameters using the excursion set model. Differences between the input and recovered parameters would indicate the mismatch between the two algorithms.

As before, we focus on $z = 7$. The mock data is generated using the photon conserving model at a resolution of $\Delta x = 2 h^{-1}$~cMpc, i.e., the data is identical to that used in \secn{sec:two_param_model}. We first check the parameter recovery for the excursion set model at the same resolution, the results are shown in \fig{fig:PC_mock_ES}. As is obvious, the recovered parameters are significantly different from the input ones. In particular, the recovered $Q_{\mathrm{HII}}^M$ is different by $\sim 10\%$ from the input value, as can be seen in \tab{tab:table_mock_pc_es}. This would then be the typical level of discrepancy between the two algorithms at the mid-stages of reionization.

It was also shown by \citet{2018MNRAS.481.3821C} that the amount of photon non-conservation becomes less when the resolution is coarser. To check the implications, we also run a MCMC with the  coarser resolution of $\Delta x = 4 h^{-1}$~cMpc. The corresponding results (red) in \fig{fig:PC_mock_ES} indicate that the recovery of $Q_{\mathrm{HII}}^M$ is much better in this case. The same conclusion can be drawn from \tab{tab:table_mock_pc_es} which shows that the recovered $Q_{\mathrm{HII}}^M$ is within $\sim 2.5\%$ of the input value. Thus, the match between excursion set and photon conserving models improves as we coarsen the resolution.

We have also run the chains for $z = 8$ and $z = 6.2$ for completeness. The  recoveries of $Q_{\mathrm{HII}}^M$ at $z=8.0$ is reasonably well (input value is with $1 \sigma$ uncertainties). This is due to that the photon number conservation is less dominant at higher redshifts when the global ionization fraction is small. On the other hand, the recoveries in the case of $z=6.2$ are worse.  

This analysis confirms that the recovery of the reionization history using 21~cm power spectra depends on the semi-numerical algorithm used for generating the ionization maps. It thus becomes important to develop some kind of consensus within the community of researchers working on semi-numerical models of reionization as to what would be the best way of identifying a physically meaningful model which remains computationally efficient.

\section{Numerical convergence of the 21~cm power spectra for PC models} 

\begin{figure*}
    \centering
    \includegraphics[width=0.99\textwidth]{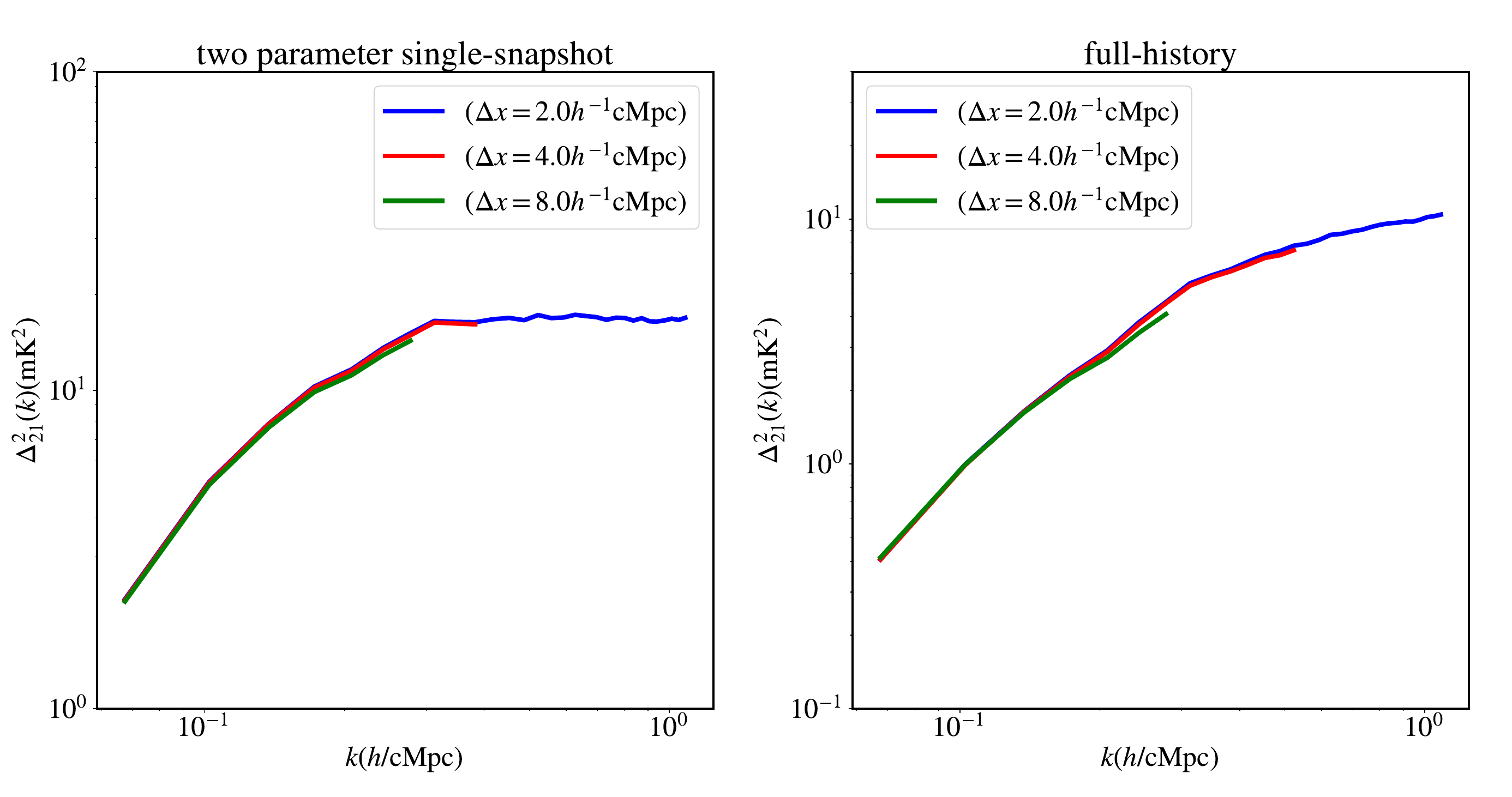}
    \caption{The 21~cm power spectra at $z = 7$ for three different grid sizes $\Delta x = 2, ~4~ \&~8~h^{-1}\mathrm{cMpc}$ used for generating the ionized maps. We show the results for the different fiducial models used in the study.}
    \label{fig:conv_ch}
\end{figure*}

We study the numerical convergence of the 21~cm power spectra obtained using our photon conserving models of reionization. In \fig{fig:conv_ch}, we show the power spectra at $z = 7$ for three different grid sizes, namely, $\Delta x = 2, ~4~ \&~8~h^{-1}\mathrm{cMpc}$ for the different fiducial models used in the study. It is evident that the power spectra at large scales convergence to within $5\%$ with respect to the resolutions used for generating the maps. The numerical convergence for the two-parameter single snapshot model was already shown by \citet{2018MNRAS.481.3821C}. It is interesting that the results hold also for the full-history models.

\bsp	
\label{lastpage}
\end{document}